\documentclass[12pt]{article}
\usepackage{graphicx}
\usepackage{epsfig}
\begin {document}
\begin{center}
{\bf FEYNMAN SCALING VIOLATION\\
DUE TO BARYON NUMBER DIFFUSION\\
IN RAPIDITY SPACE}

\vspace{.2cm}

G.H. Arakelyan$^*$, C. Merino$^{**}$, C. Pajares$^{**}$,
and Yu.M. Shabelski$^{***}$ \\

\vspace{.5cm}
$^*$ A.I.Alikhanyan National Science Laboratory (Yerevan Physics
Institute)\\
Yervan, Armenia\\
E-mail: argev@mail.yerphi.am

\vspace{.2cm}

$^{**}$ Departamento de F\'\i sica de Part\'\i culas and
\\
Instituto Galego de F\'\i sica de Altas Enerx\'\i as (IGFAE)\\
Universidade de Santiago de Compostela\\
Santiago de Compostela, Galiza, Spain\\
E-mail: merino@fpaxp1.usc.es \\
E-mail: pajares@fpaxp1.usc.es \\

\vspace{.2cm}

$^{**}$ Petersburg Nuclear Physics Institute\\
St.Petersburg, Russia\\
E-mail: shabelsk@thd.pnpi.spb.ru
\vskip 0.9 truecm

\vspace{1.2cm}

A b s t r a c t
\end{center}

A significant asymmetry in baryon/antibaryon yields in the central
region of high energy collisions is observed when the initial state
has non-zero baryon charge. This asymmetry is connected with the
possibility of a baryon charge diffusion in rapidity space.
Evidently, such a diffusion should decrease the baryon charge in the
fragmentation region leading to the corresponding decrease of the
multiplicity of leading baryons. As a result, a new mechanism
for Feynman scaling violation in the fragmentation region is obtained.
We present the quantitative predictions for the Feynman scaling violation
at LHC energies and even at higher energies that can be important for
cosmic ray physics.

\vskip 1.5cm

PACS. 25.75.Dw Particle and resonance production

\newpage

\section{Introduction}


The problem of Feynman scaling violation has evident both theoretical and
practical interest. In particular, this question is very
important \cite{Abr,Abb} for cosmic ray physics, where the difference
from the primary radiation to the events registrated on the ground or
mountain level is determined by the multiple interactions of
the so-called leading particles (mainly baryons) in the atmosphere.

Despite the lack of direct measurements of Feynman scaling violation for
secondary baryon spectra in nucleon-nucleon collisions at energies higher
than those of ISR, some experimental information from cosmic ray
experiments seems to confirm \cite{ELF,Ver,KKh} the presence of significant
Feynman scaling violation effects. Now the LHCf Collaboration has started the
search \cite{LHCf1,LHCf2} of Feynman scaling violation effects for the
spectra of photons ($\pi^0$) and neutrons in the fragmentation region
at LHC energies.

In principale, the violation of Feynman scaling in the fragmentation region
should exist due to the energy conservation, since the spectra of
charged particles increase in the central region. However, no quantitative
predictions can be made without some model of the particle production.
Actually, even a small decrease of the spectrum of some secondaries
in a narrow region near $x_F=1$ may be enough to satisfy the energy
conservation law.

The predictions of the Additive Quark Model \cite{ASS} for Feynman
scaling violation in the fragmentation region were considered in \cite{ABS}.
In this model, the scaling violation effects are connected with
the increase of the interaction cross sections, which leads to the decrease
in the number of quark-spectators which form the spectra of fast
secondaries. However, to make a description of the energy dependences of
the spectra as a function of $x_F$ additional assumptions and parameters
are needed.

The Quark-Gluon String Model (QGSM)~\cite{KTM} has some important
advantages in this respect, since it is an analytical model and it allows
the calculation of the spectra of secondaries at different initial energies
in the whole $x_F$ region. The QGSM is based on Dual Topological
Unitarization (DTU), Regge phenomenology, and nonperturbative notions
of QCD. This model is successfully used for the description of
multiparticle production processes in hadron-hadron
\cite{KaPi,Sh,Ans,ACKS}, hadron-nucleus \cite{KTMS,Sh1}, and
nucleus-nucleus \cite{JDDS} collisions.

In the QGSM high energy interactions are considered as proceeding via the
exchange of one or several Pomerons, and all elastic and inelastic processes
result from cutting through or between Pomerons \cite{AGK}. Inclusive
spectra of hadrons are related to the corresponding fragmentation functions
of quarks and diquarks, which are constructed using the Reggeon counting
rules \cite{Kai}. The quantitative predictions of the QGSM depend on
several parameters which were fixed by comparison of the theoretical calculations
to the experimental data obtained at fixed target energies. The first
experimental data obtained at LHC show \cite{MPRS} that the model
predictions are in reasonable agreement with the data.

In the frame of QGSM several reasons of Feynman scaling violation
in the fragmentation region exist. The first one is the increase of the
average number of exchanged Pomerons with the energy, which leads to the
corresponding increase of the yields of hadron secondaries in the central
region and to their decrease in the fragmentation region. This effect was
considered in~\cite{Sh2,Sh3}.

In the case of nuclear (air) targets, the growth of the $hN$ cross section
with energy leads to the increase of the average number of fast hadron
inelastic collisions inside the nucleus. Thus, the average number of Pomerons
is additionally increased, resulting in a stronger Feynman scaling
violation~\cite{Sh2,Sh3}.

In reference~\cite{EKS} these predictions were taken into account to calculate
the penetration of fast hadrons into the atmosphere, leading to a better
description of the cosmic ray experimental data.

The differences in the yields of baryons and antibaryons produced in the
central (midrapidity) region of high energy $pp$ interactions
\cite{ACKS,MPRS,BS,AMPS,MRS1,AKMS,MPS1} are significant. Evidently,
the appearance of the positive baryon charge in the central region of $pp$
collisions should be compensated by the decrease of the baryon multiplicities
in the fragmentation region that leads to an additional reason for Feynman
scaling violation.

In the present paper we consider the effects of Feynman scaling violation
in $pp$ collisions through large distance baryon diffusion in rapidity space.
The role of the nuclear factor for air nuclei should be
of the same magnitude as for that presented in \cite{Sh2,Sh3}.

\section{Baryon/antibaryon asymmetry in the QGSM}

The Quark-Gluon String Model (QGSM) \cite{KTM,KaPi,Sh} allows us to make
quantitative predictions of different features of multiparticle production.
In QGSM the inclusive spectrum of a secondary hadron $h$ is determined by the
convolution of the diquark, valence quark, and sea quark distributions
$u(x,n)$ in the incident particles with the fragmentation functions $G^h(z)$
of quarks and diquarks into the secondary hadron $h$. These distributions,
as well as the fragmentation functions, are constructed using the Reggeon
counting rules \cite{Kai}. The details of the model are presented in
\cite{KTM,KaPi,Sh,ACKS}. The Pomeron parameters were taken from \cite{Sh}.

In the string models, baryons are considered as configurations
consisting of three connected strings (related to three valence quarks)
called string junction (SJ) \cite{Artru,IOT,RV,Khar}. Such a baryon
structure is supported by lattice calculations \cite{latt}. This picture
leads to some general phenomenological predictions. In the
case of inclusive reactions the baryon number transfer to large rapidity
distances in hadron-nucleon and hadron-nucleus reactions can be explained
\cite{ACKS,BS,AMS,Olga,MRS} by SJ diffusion.

The production of a baryon-antibaryon pair in the central region usually
occurs via $SJ$-$\overline{SJ}$ (SJ has upper color indices whereas
$\overline{SJ}$ has lower indices) pair production, which then
combines with sea quarks and sea antiquarks into a $B\overline{B}$
pair \cite{RV,VGW}, as it is shown in Fig.~1a.
\begin{figure}[htb]
\centering
\includegraphics[width=.45\hsize]{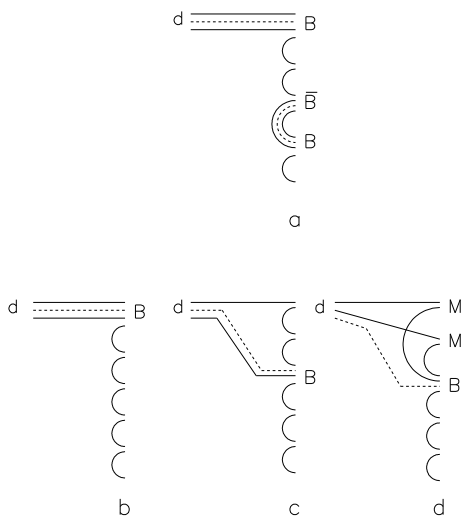}
\caption{\footnotesize
QGSM diagrams describing secondary baryon $B$ production by diquark $d$.
(a) Central production of $\overline{B}B$ pair. Single $B$ production
in the processes of diquark fragmentation: (b) initial SJ together with two
valence quarks and one sea quark, (c) initial SJ together with one valence
quark and two sea quarks, and (d) initial SJ together with three sea quarks.
Quarks are shown by solid curves and SJ by dashed curves.}
\end{figure}

In the processes with incident baryons, e.g. in $pp$ collisions, another
possibility to produce a secondary baryon in the central region exists.
This possibility is the diffusion in rapidity space of any SJ existing
in the initial state and it can lead to significant differences in the yields
of baryons and antibaryons in the midrapidity region even at rather high
energies~\cite{ACKS}. The most important experimental fact
in favour of this process is the rather large asymmetry in
$\Omega$ and $\overline{\Omega}$ baryon production in high energy $\pi^-p$
interactions~\cite{ait}.

The theoretical quantitative description of the baryon number transfer
via SJ mechanism was suggested in the 90's and used to predict~\cite{KP1}
the $p/\overline{p}$ asymmetry at HERA energies.

In order to obtain the net baryon charge we consider, following ref.
\cite{ACKS} three different possibilities. The first one is the fragmentation
of the diquark giving rise to a leading baryon (Fig.~1b). A second possibility
is to produce a leading meson in the first break-up of the string and a baryon
in a subsequent break-up (Fig.~1c). In these two first cases the baryon
number transfer is possible only for short distances in rapidity. In the
third case, shown in Fig.~1d, both initial valence quarks recombine with
sea antiquarks into mesons $M$ while a secondary baryon is formed by the
SJ together with three sea quarks.

The fragmentation functions for the secondary baryon $B$ production
corresponding to the three processes shown in Figs.~1b, 1c, and 1d, can
be written as follows~\cite{ACKS}:
\begin{eqnarray}
G^B_{qq}(z) &=& a_N\cdot v^B_{qq} \cdot z^{2.5} \;, \\
G^B_{qs}(z) &=& a_N\cdot v^B_{qs} \cdot z^2\cdot (1-z) \;, \\
G^B_{ss}(z) &=& a_N\cdot\varepsilon\cdot v^B_{ss} \cdot z^{1 - \alpha_{SJ}}
\cdot (1-z)^2 \;,
\end{eqnarray}
where $a_N$ is the normalization parameter, and $v^B_{qq}$, $v^B_{qs}$,
$v^B_{ss}$ are the relative probabilities for different baryons production
that can be found by simple quark combinatorics \cite{AnSh,CS}. Their
numerical values for different secondary baryons were presented in
\cite{AKMS}.

The first two processes shown in Figs.~1b and 1c, Eqs.~(1) and (2),
determine the spectra of leading baryons in the fragmentation region. The
third contribution shown in Fig.~1d, Eq.~(3), is essential if the value of
the intercept of the SJ exchange Regge-trajectory, $\alpha_{SJ}$, is not
too small. In QGSM the weight of this third contribution is determined by
the coefficient $\varepsilon$ which fixes the small probability for such
a baryon number transfer to occur.

At high energies, the SJ contribution to the inclusive cross section
of secondary baryon production at large rapidity distance $\Delta y$ from
the incident nucleon can be estimated as
\begin{equation}
(1/\sigma)d\sigma^B/dy \sim a_B\cdot\varepsilon\cdot
e^{(1 - \alpha_{SJ})\cdot\Delta y} \;,
\end{equation}
where $a_B = a_N\cdot v^B_{ss}$. The baryon charge transferred to large
rapidity distances can be determined by integration of Eq.~(4), so it is
of the order of
\begin{equation}
\langle n_B \rangle \sim a_B\cdot\frac{\varepsilon}{(1 - \alpha_{SJ})} \; ,
\end{equation}
and so, only the left part of the initial baryon charge can be used for
the production of the leading baryons.

To obtain the QGSM predictions for the spectra of leading baryons we use
the standard expressions of Reggeon theory and QGSM \cite{KTM,ACKS}.

Though currently the value of $\alpha_{SJ} = 0.5$ seems more
plausible~\cite{ALICE}, the value of $\alpha_{SJ} = 0.9$ can not be
excluded~\cite{LHCb,Conf}. Thus, in this paper we present the calculation
obtained with these two values of $\alpha_{SJ}$,
and also without any SJ contribution ($\varepsilon = 0.$).

\section{Spectra of baryons in $ep$ collisions}

The only very high energy reaction where the leading secondary baryons
were measured is $ep$ collisions at HERA. Here, both $ep \to epX$ (i.e.
$\gamma p \to pX$) \cite{ZEU1,ZEU2} and $ep \to enX$ (i.e. $\gamma p \to nX$)
\cite{ZEU3} were investigated at $W \sim$ 200 GeV. Due to the vector
dominance principle, the inelastic $\gamma p$ interaction can be considered
as a superposition of $\rho^0 p$ and $\omega p$\footnote{As we do not
consider the yields of strange secondaries, the numerically small contribution
of $s\overline{s}$ pairs ($\phi$-meson) is not important here.} intercations.
In the frame of the QGSM these two interactions are equivalent to the sum of
$(1/2)\pi^+ p + (1/2)\pi^- p$ collisions.

In Fig.~2 the experimental data for $\gamma p \to pX$ \cite{ZEU1,ZEU2} and for
$\gamma p \to nX$ \cite{ZEU3} are compared with the QGSM
calculations. Here the experimental data are presented as depending
on $x_L$ which is very close to $x_F$ if $p_T$ is small.
The comparison of the $ep \to epX$ data with the QGSM calculations is shown
in the upper panel of Fig.~2, while the data of the $ep \to enX$ reaction
at low $Q^2$ and high $Q^2$ are presented in the middle and lower panels of
Fig.~2, respectively.

The solid curves correspond to the calculations with $\alpha_{SJ} = 0.9$,
dashed curves to those with $\alpha_{SJ} = 0.5$, and dotted curves are
calculated without SJ contributions ($\varepsilon = 0.$). Sice the energy
dependences of all curves are weak, they all are calculated at $W = 200$ GeV.
The difference between the curves is small. All experimental points are
obtained at rather small values of the transverse momenta of the secondary
baryons, so they lie below the theoretical curves that are calculated for
the spectra integrated over $p_T$.

Usually, the spectra of neutrons produced in DIS at large $x_L$
($x_L \geq 0.7$) and small $p_T$ are described in the framework of the
one-pion-exchange approach \cite{KKMR,KPSS}. This approach leads to a more
detailed description of the data, for example it allows one to calculate
the $x_L$-spectra in different $p_T$ regions. However, our QGSM description
based on quark and diquark fragmentation picture gives reasonable
results for $x_L \leq 0.8$.

\newpage
\begin{figure}[htb]
\vskip -1.75cm
\centering
\vskip -1.cm
\includegraphics[width=.6\hsize]{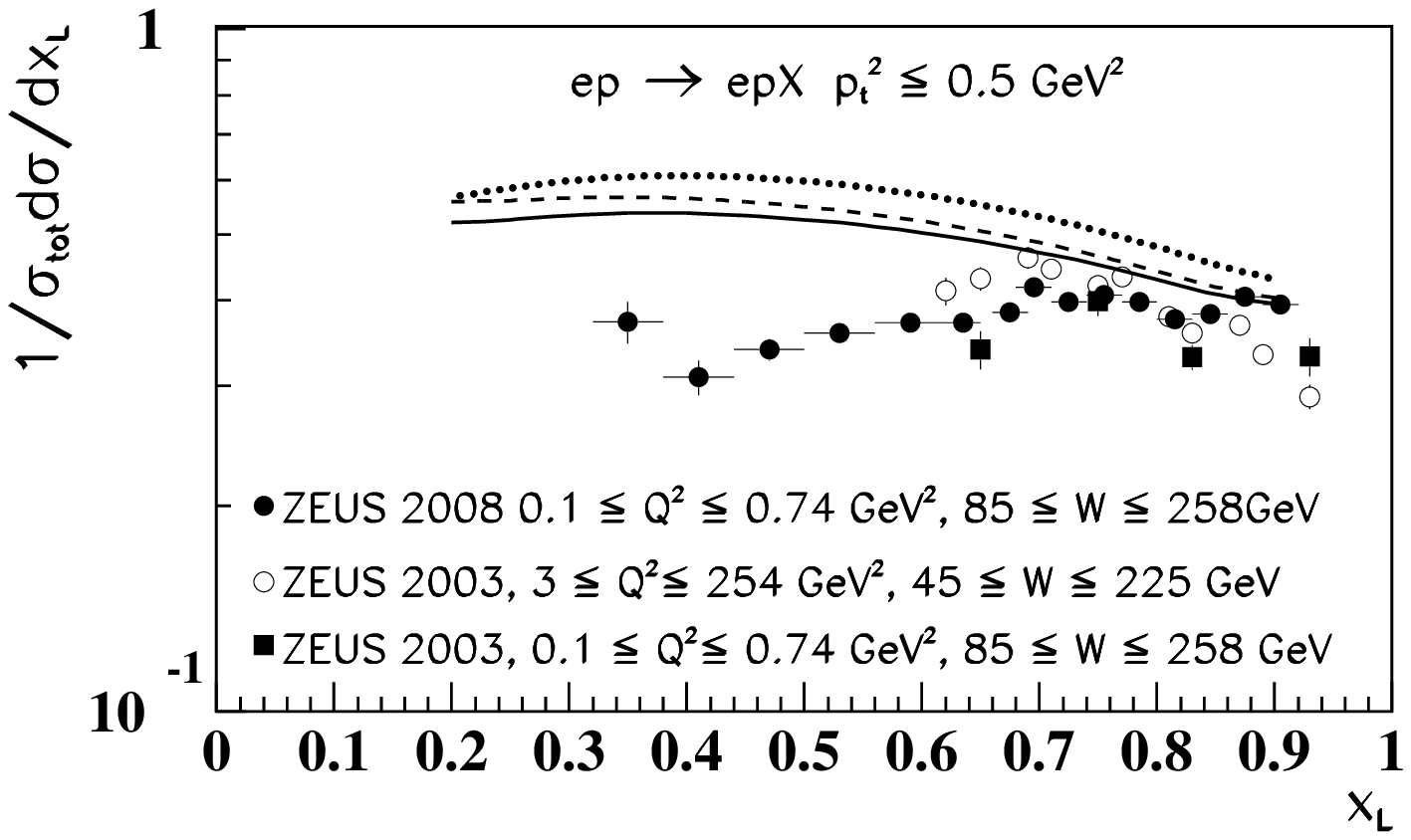}
\vskip -1.5cm
\includegraphics[width=.6\hsize]{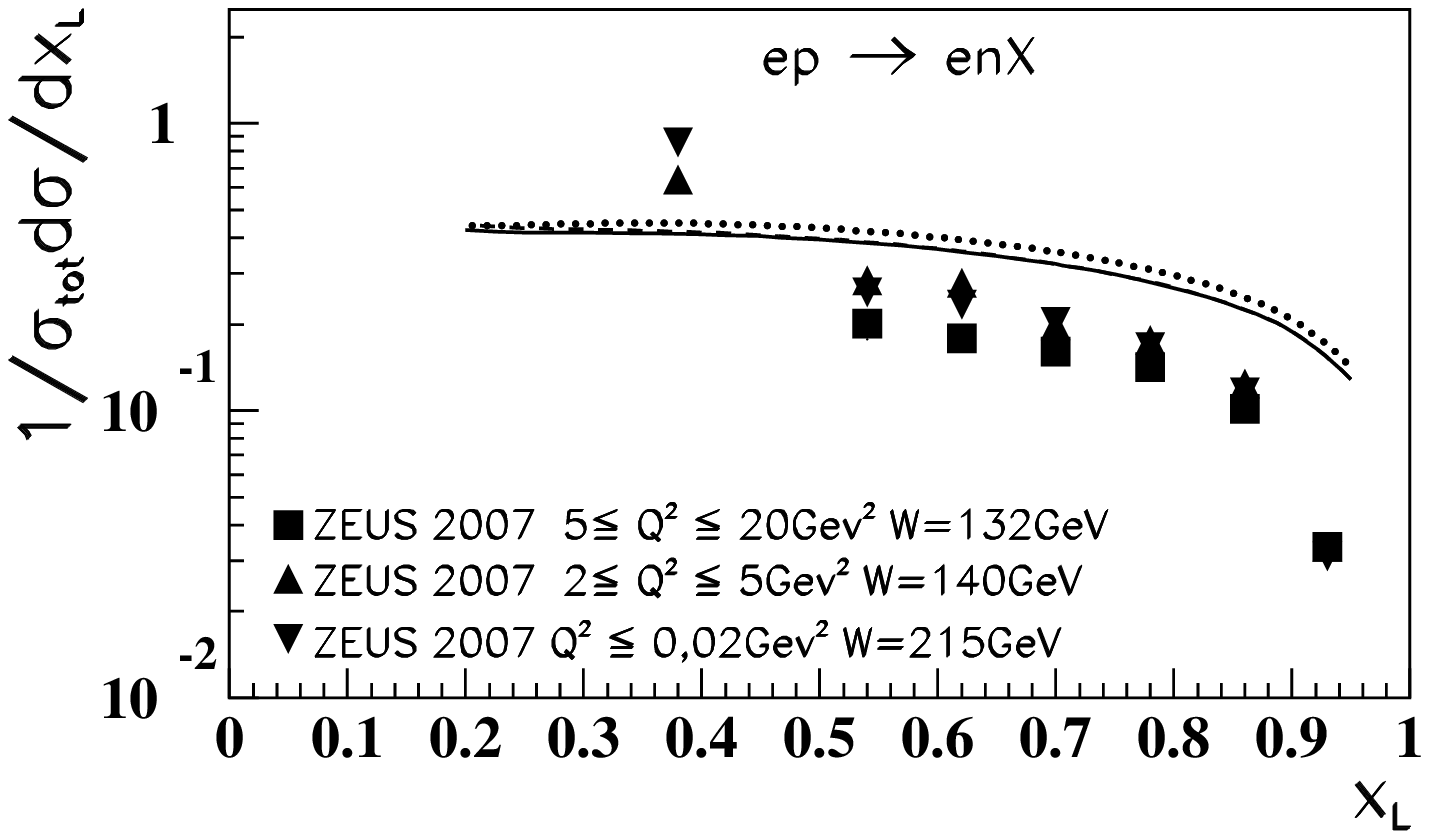}
\vskip -1.5cm
\includegraphics[width=.6\hsize]{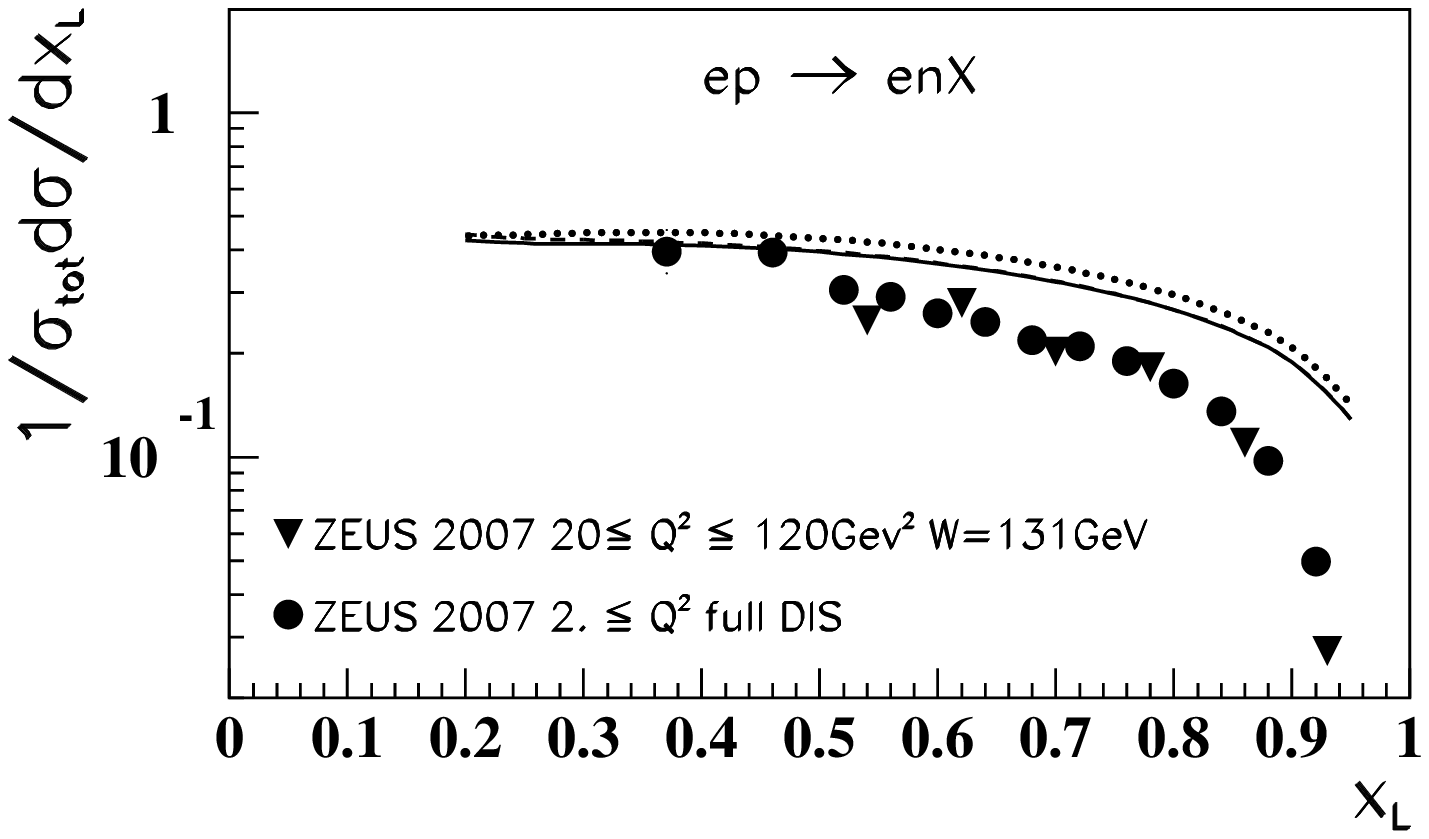}
\vskip -.9cm
\caption{\footnotesize
The QGSM predictions for the spectra of secondary protons (upper panel)
and neutrons (middle and lower panels) produced in $\gamma p$ collisions,
together with the experimental data of references \cite{ZEU1,ZEU2} and \cite{ZEU3}.
Solid curves correspond to the value $\alpha_{SJ} = 0.9$, dashed curves to
the value $\alpha_{SJ} = 0.5$, and dotted curves are calculated
without SJ contribution ($\varepsilon = 0.$).}
\end{figure}

\section{Predictions for the spectra of baryons in $pp$ collisions}

The QGSM predictions for the inclusive spectra of secondary protons,
neutrons, and $\Lambda$ at energies $\sqrt{s} = 200$ GeV, 900 GeV,
7 TeV, and 100 TeV are presented in Figs.~3, 4, and 5. In all the
calculations we have accounted for the exact conservation of the
baryon charge.

\begin{figure}[htb]
\centering
\vskip -1.75cm
\includegraphics[width=.49\hsize]{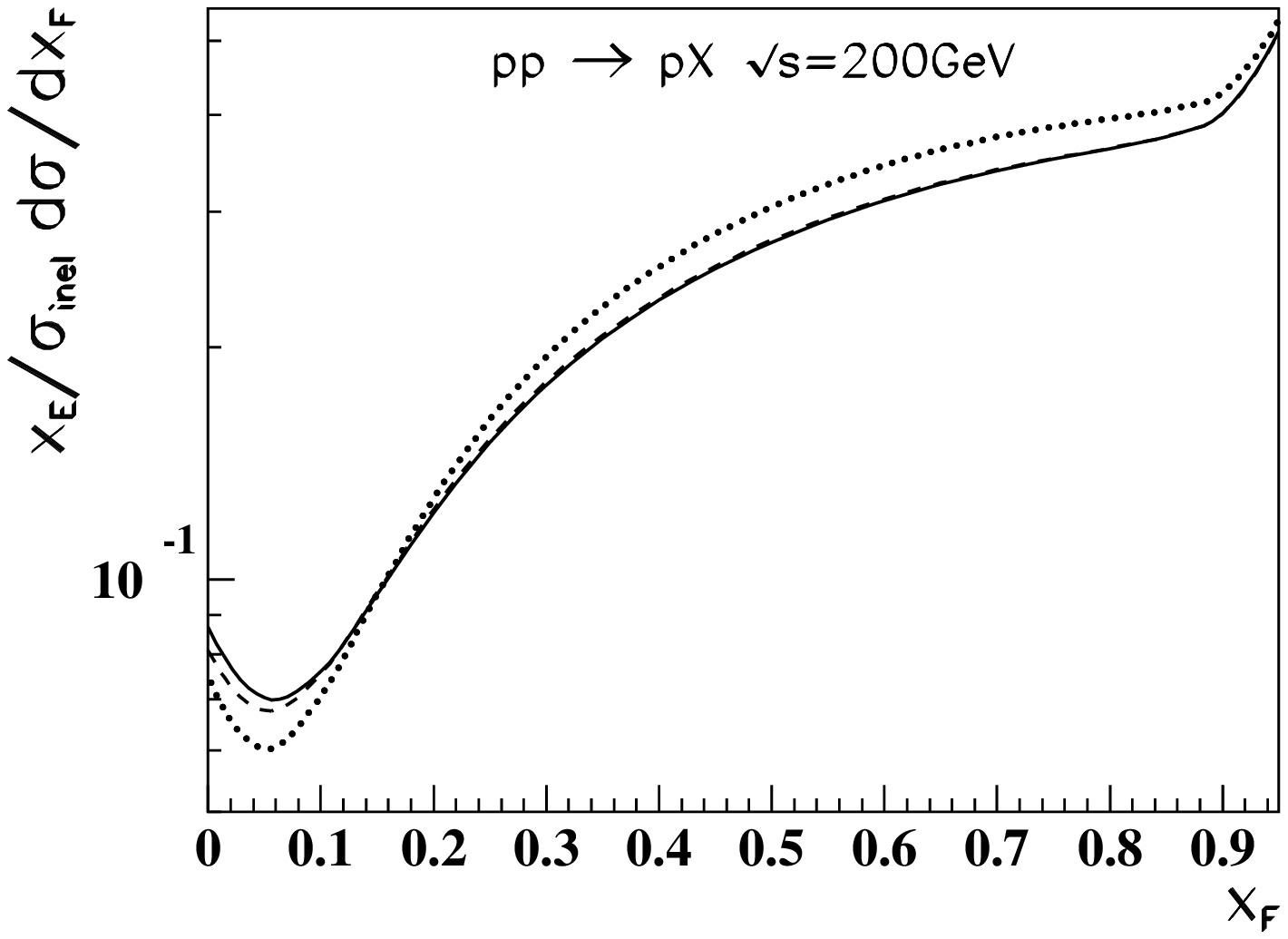}
\includegraphics[width=.49\hsize]{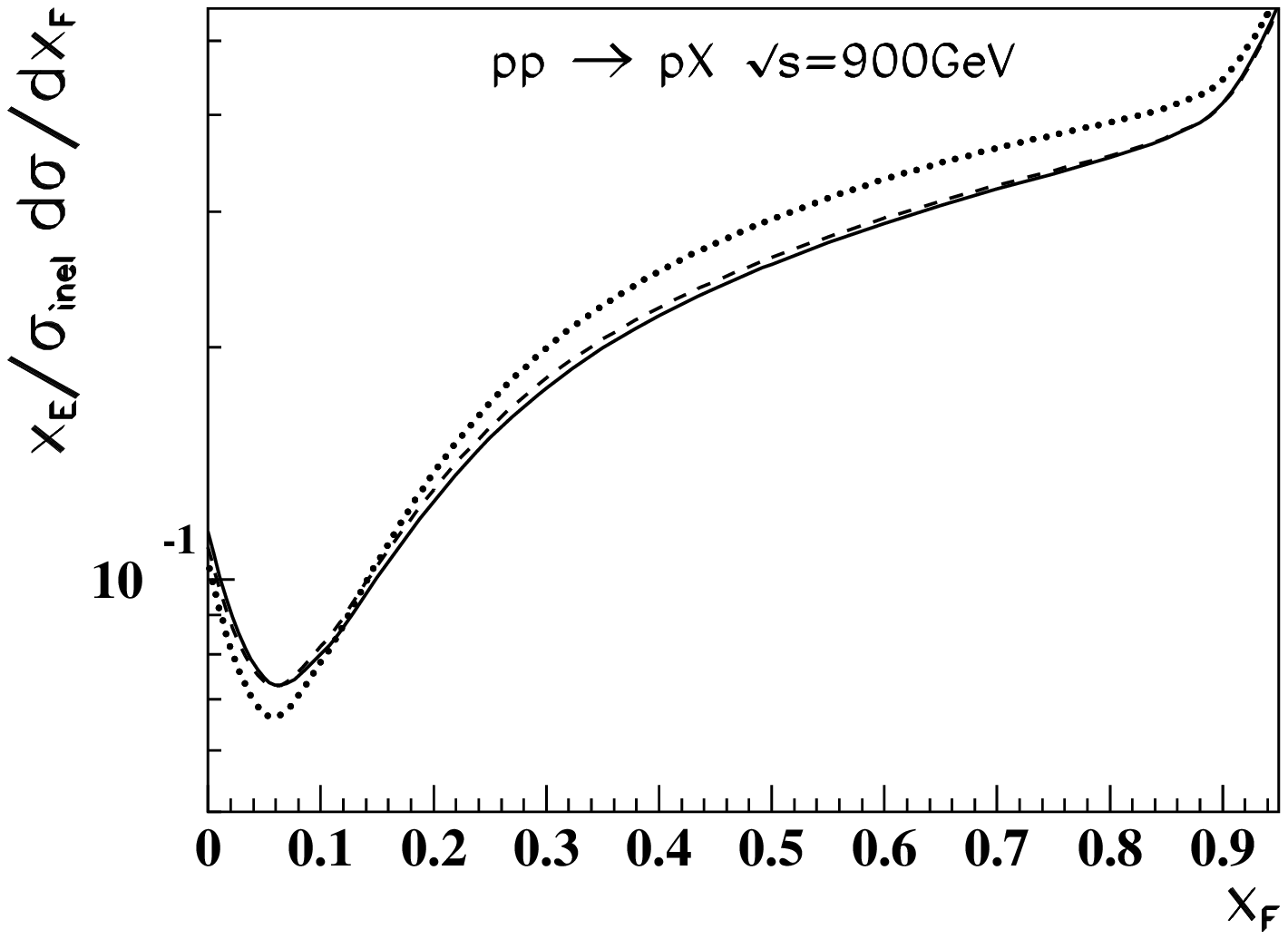}
\vskip -.6cm
\includegraphics[width=.49\hsize]{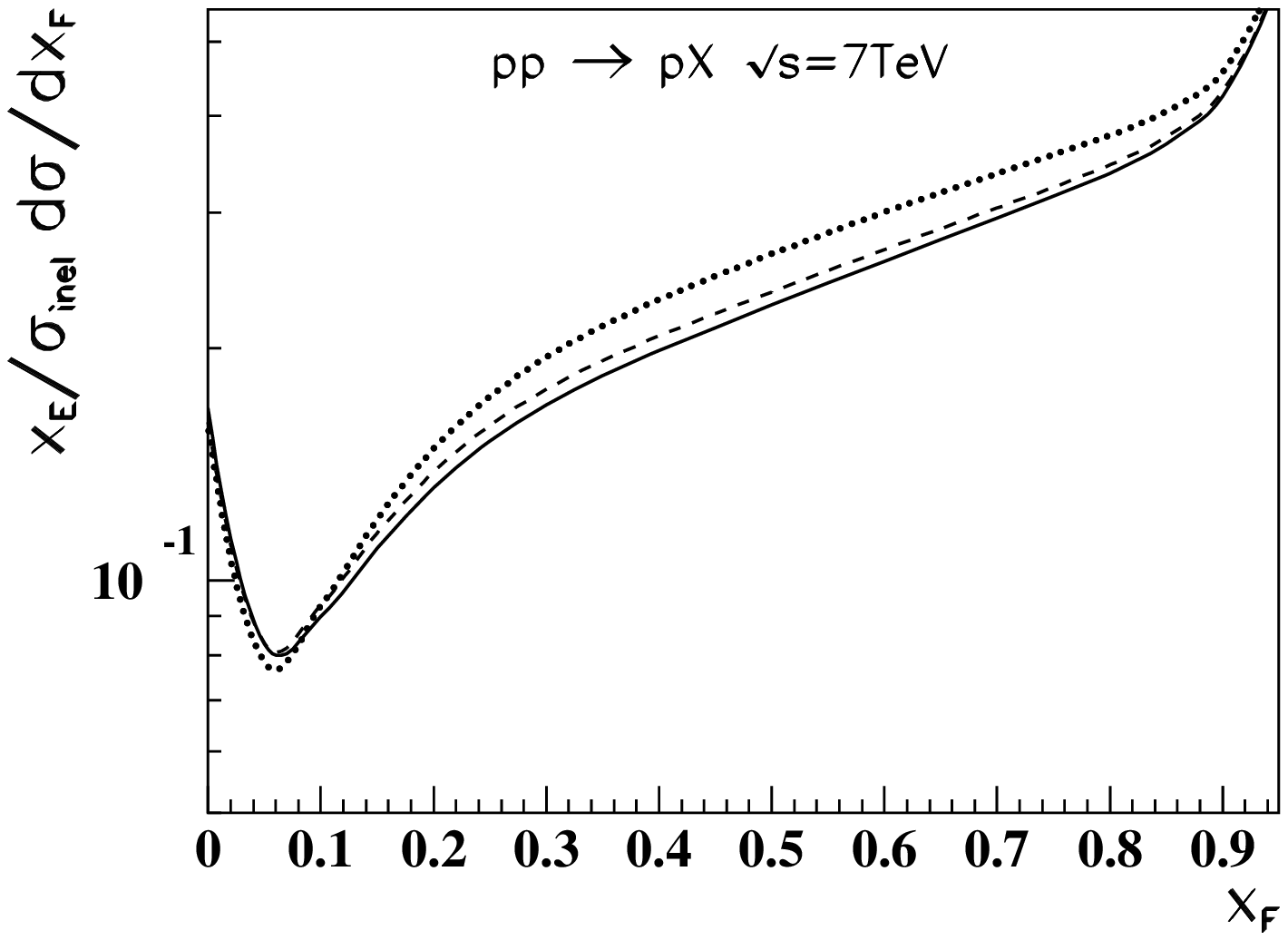}
\includegraphics[width=.49\hsize]{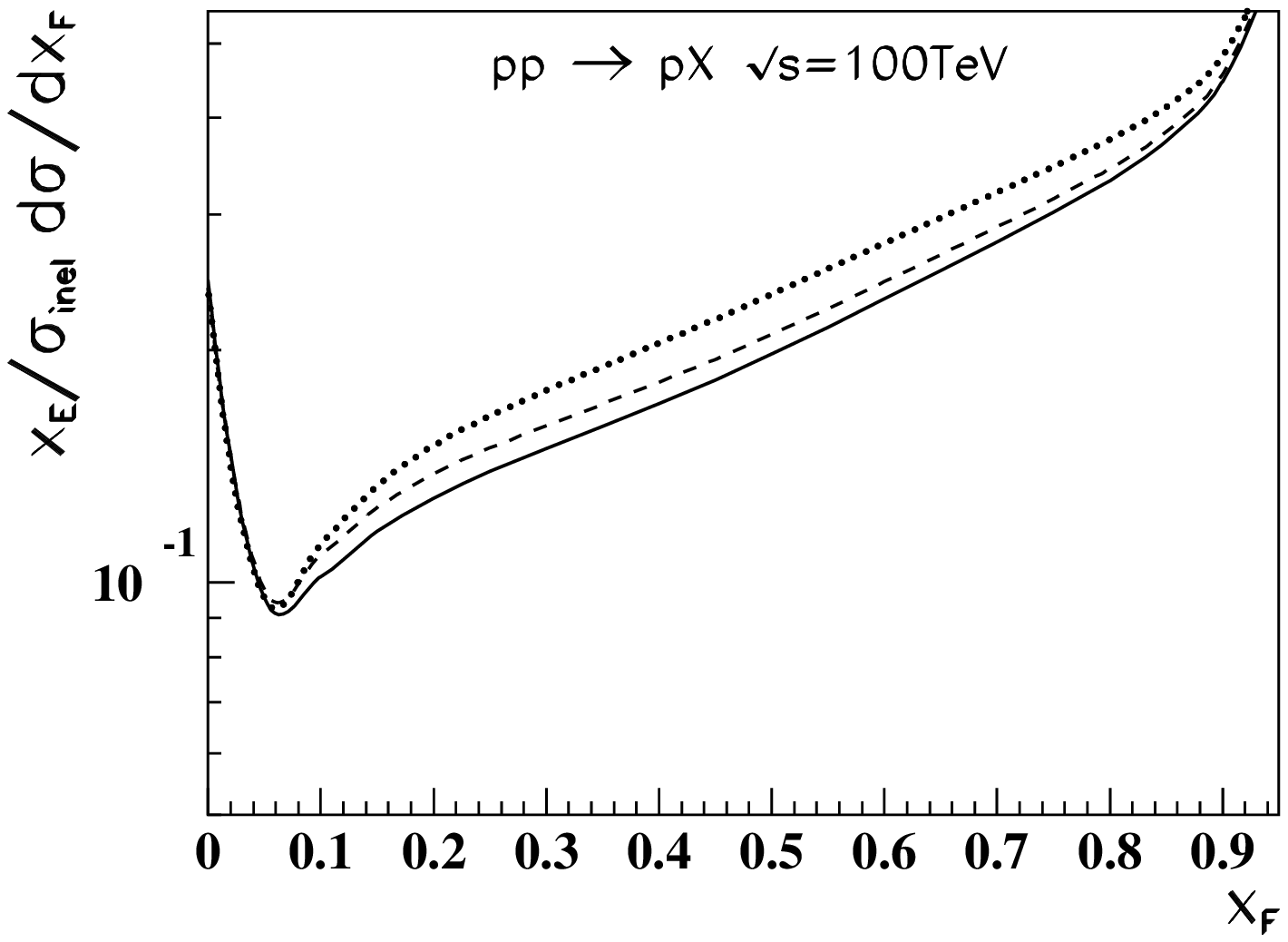}
\vskip -.6cm
\caption{\footnotesize
The QGSM predictions for the spectra of secondary protons
at energies $\sqrt{s} = 200$ GeV (up left panel), $\sqrt{s} = 900$ GeV
(up right panel), $\sqrt{s} = 7$ TeV (down left panel), and
$\sqrt{s} = 100$ TeV (down right panel). Solid curves correspond to the
value $\alpha_{SJ} = 0.9$, dashed curves to the value $\alpha_{SJ} = 0.5$,
and dotted curves are calculated without SJ contribution ($\varepsilon = 0.$).}
\end{figure}

In all Figures 3, 4, and 5 one can see the peaks at very small $x_F$, that are
connected to the contribution of the pair $B\overline{B}$ production via the
mechanism shown in Fig.~1a. The increase of the spectra with $x_F$ is connected
to the contributions of the processes shown in Figs.~1b and 1c. The
difference between both the solid and dashed curves to the dotted curves show
the effect of baryon number transfer to small $x_F$ region due to the
diagram Fig.~1d. An important result is that the Feynman scaling violation
is more sensitive to the fact of the inclusion of the baryon number diffusion
than to the exact value of $\alpha_{SJ}$.
\begin{figure}[htb]
\centering
\vskip -1.75cm
\includegraphics[width=.49\hsize]{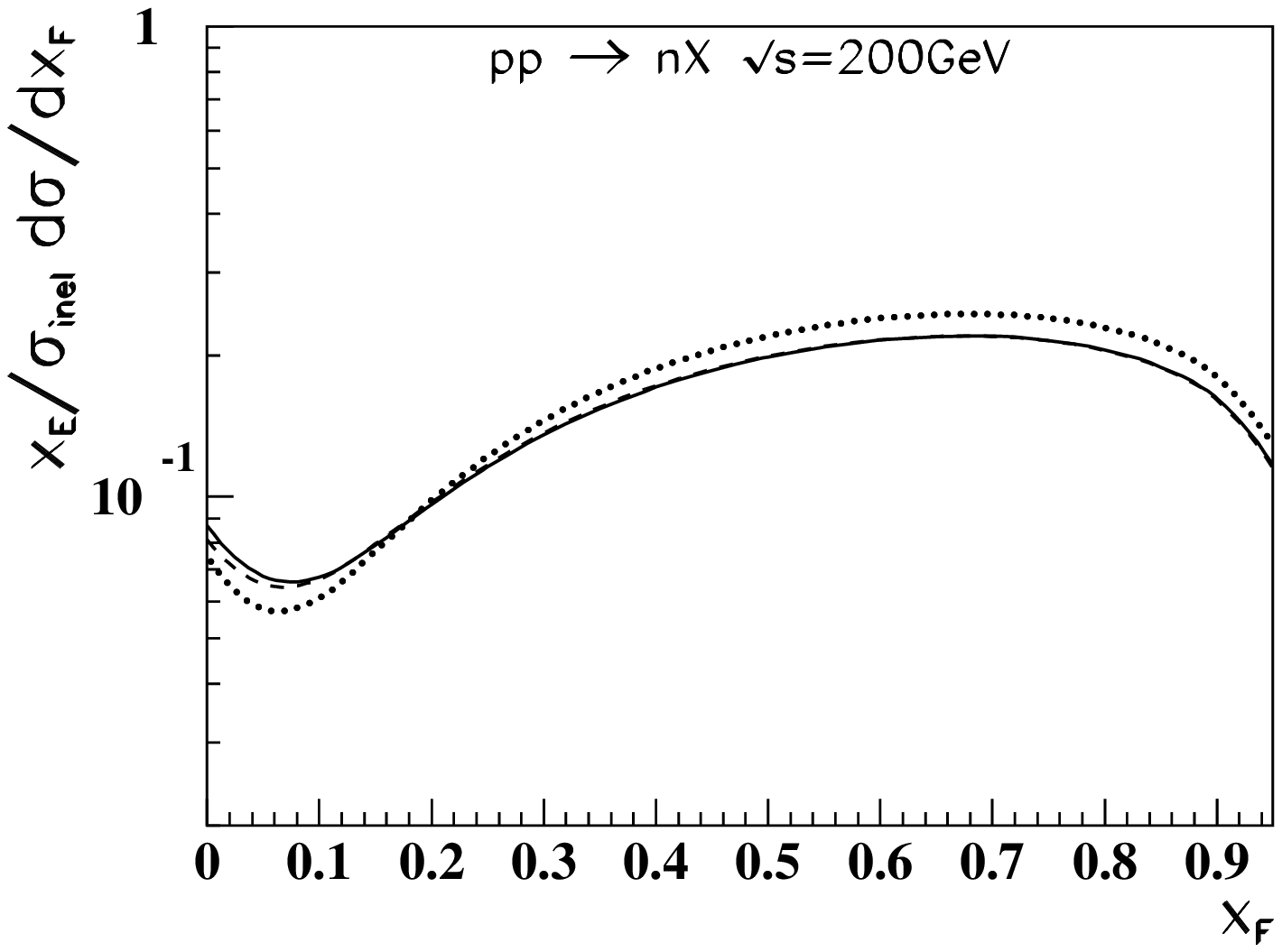}
\includegraphics[width=.49\hsize]{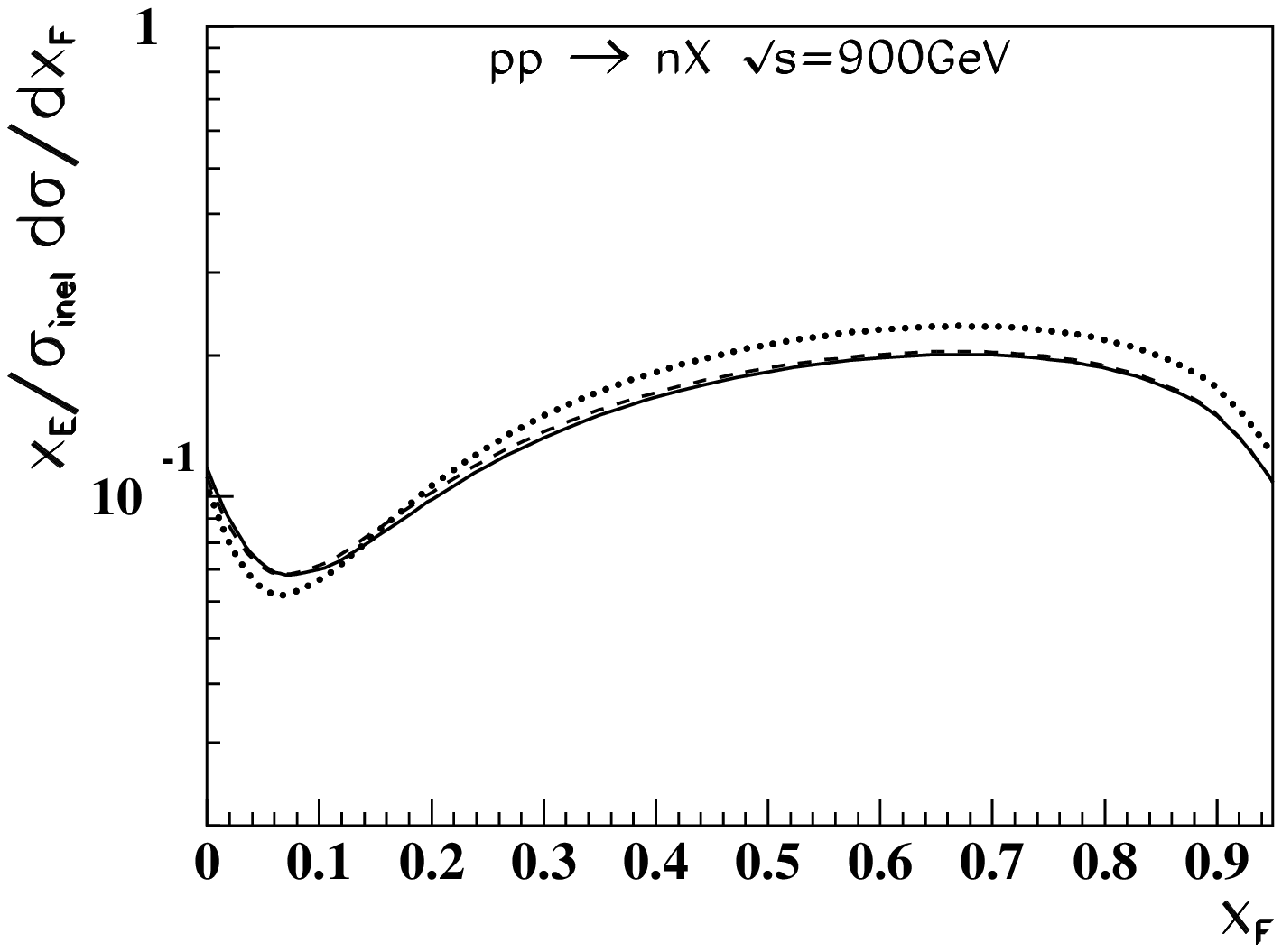}
\vskip -.6cm
\includegraphics[width=.49\hsize]{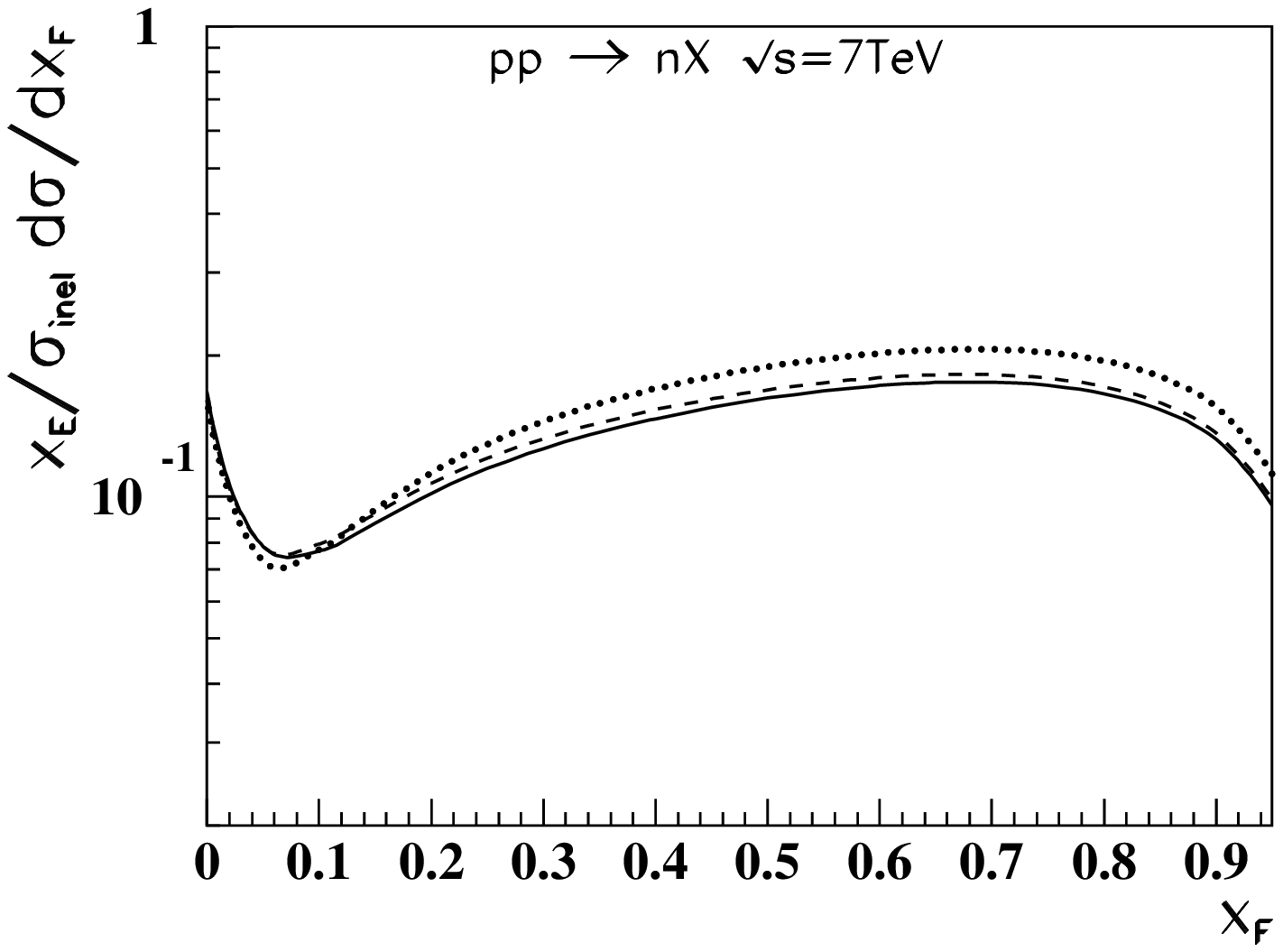}
\includegraphics[width=.49\hsize]{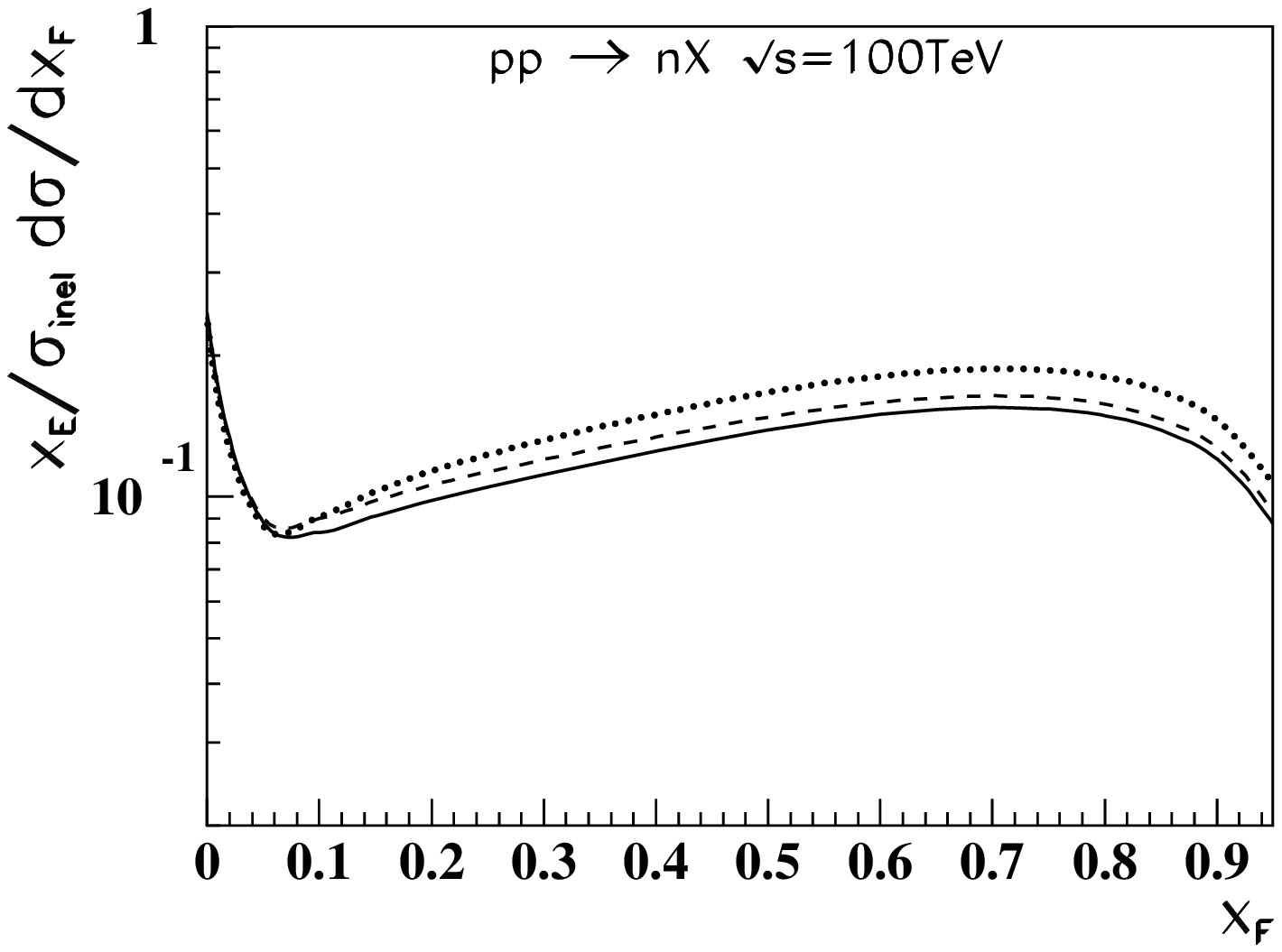}
\vskip -.6cm

\caption{\footnotesize
The QGSM predictions for the spectra of secondary neutrons at
energies $\sqrt{s} = 200$ GeV (up left panel), $\sqrt{s} = 900$ GeV
(up right panel), $\sqrt{s} = 7$ TeV (down left panel), and
$\sqrt{s} = 100$ TeV (down right panel). Solid curves correspond to the
value $\alpha_{SJ} = 0.9$, dashed curves to the value $\alpha_{SJ} = 0.5$,
and dotted curves are calculated without SJ contribution ($\varepsilon = 0.$).}
\end{figure}

One can see that in the case of secondary protons, shown in Fig.~3, the
differences in the fragmentation region between the two calculations with SJ
contribution, with values $\alpha_{SJ} = 0.9$ and $\alpha_{SJ} = 0.5$, are
very small at energies $\sqrt{s} \leq 1$ TeV, while they increase up to
5-7\% at higher energies. In the case of the calculations with and without
SJ contribution the difference in the spectra in the region $x_F = 0.5 - 0.8$
is of about 7-10\% for $\sqrt{s} = 200$ GeV and 900 GeV, and it becomes of
about 15-20\% at $\sqrt{s} = 7$ TeV and 100 TeV. So, by accounting for the possibility
of baryon charge diffusion in rapidity space one gets one additional contribution to
the Fyenman scaling violation.

Our predictions for the spectra of secondary neutrons are presented in
Fig.~4. Their behaviour with respect to the accounting of the SJ contribution
and to the effect of the Feynman scaling violation are similar to those of
secondary protons.
\begin{figure}[htb]
\centering
\vskip -1.75cm
\includegraphics[width=.49\hsize]{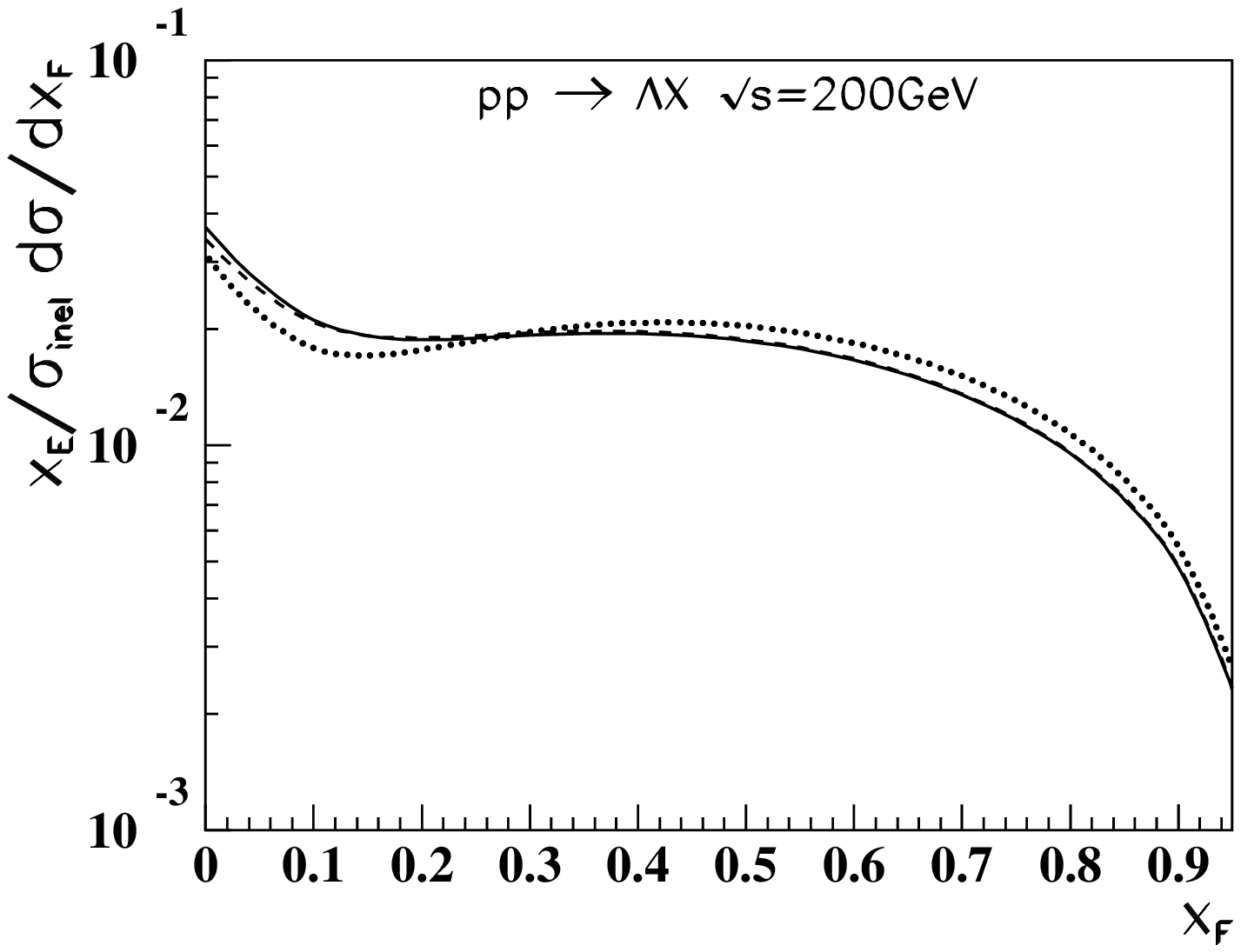}
\includegraphics[width=.49\hsize]{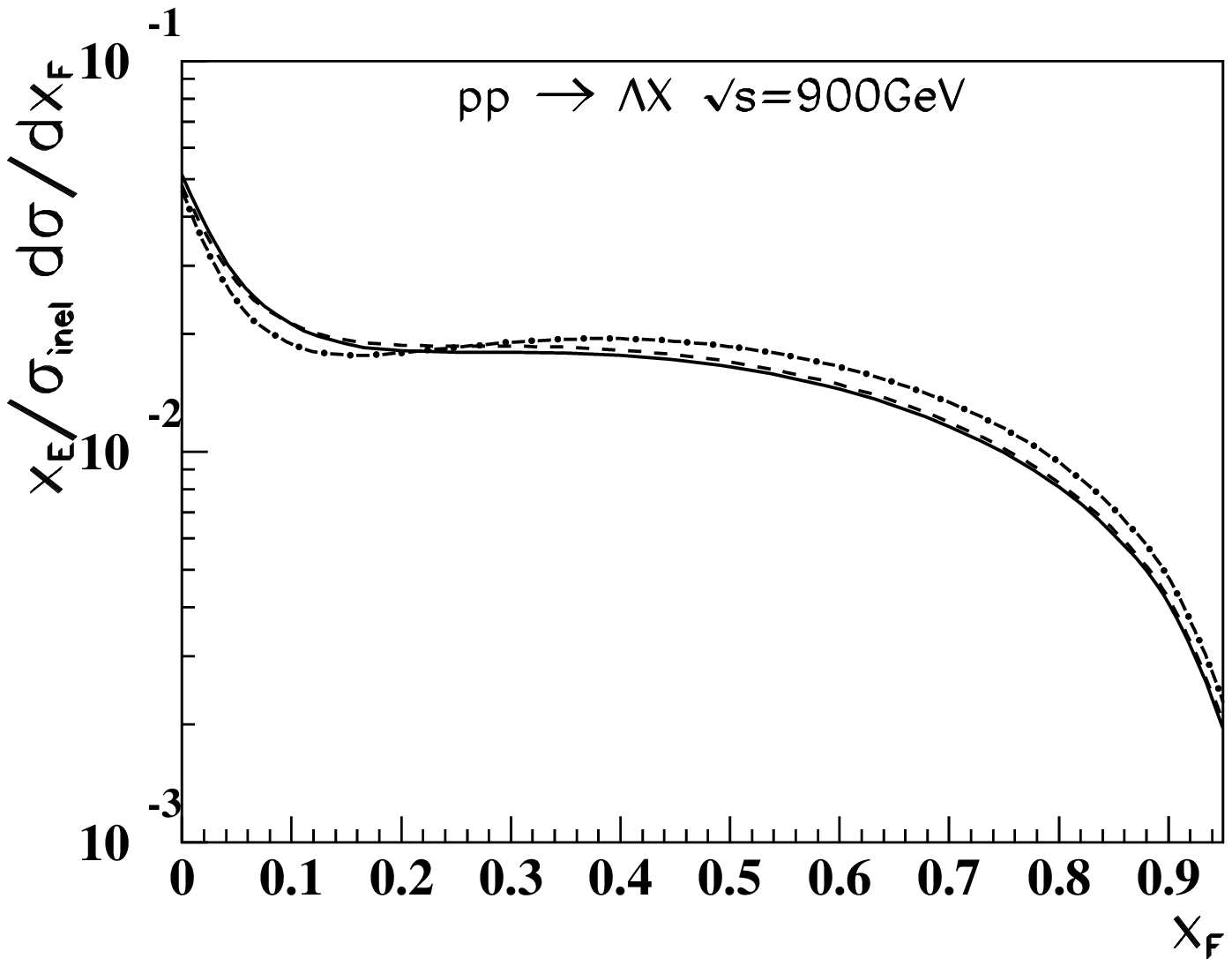}
\vskip -.6cm
\includegraphics[width=.49\hsize]{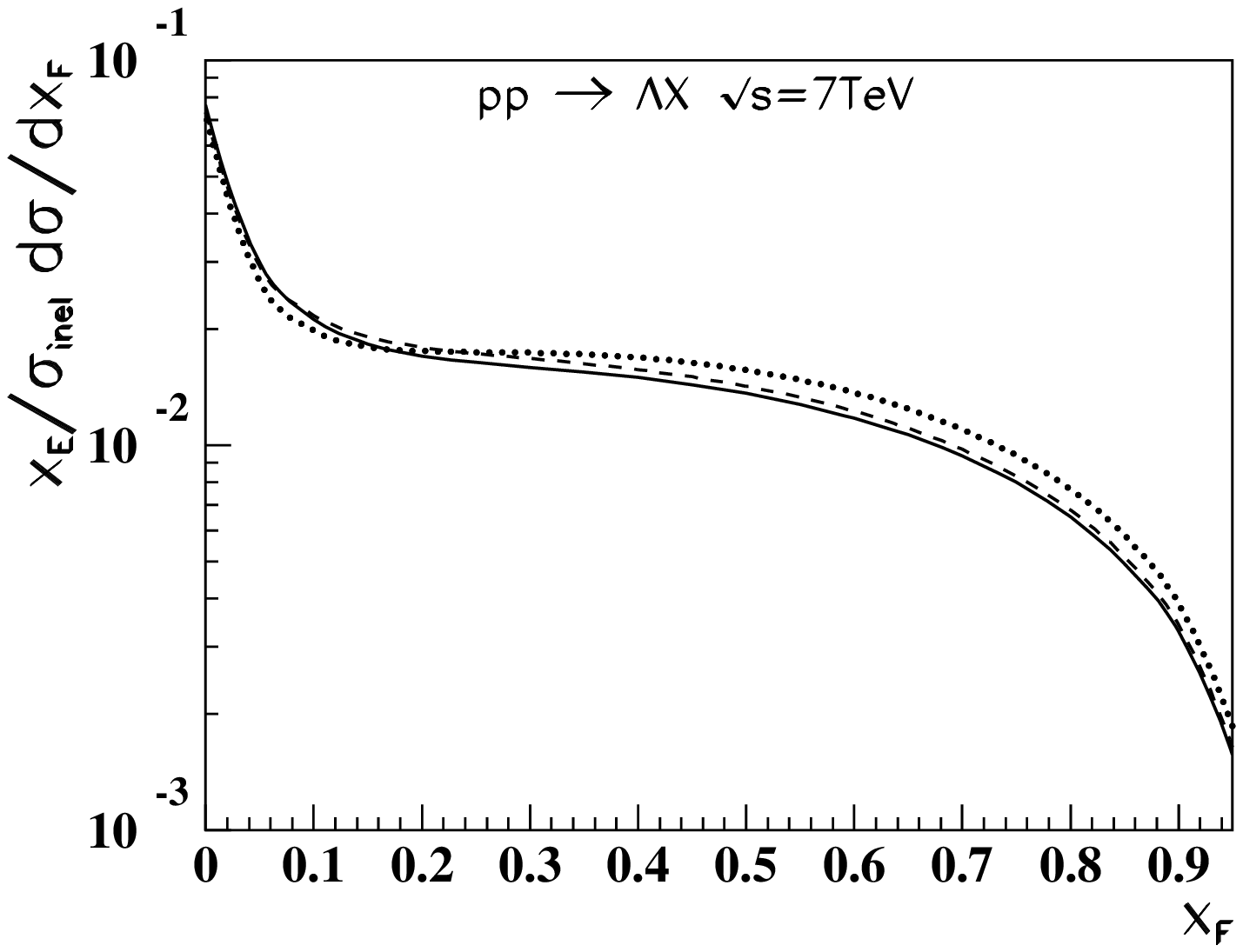}
\includegraphics[width=.49\hsize]{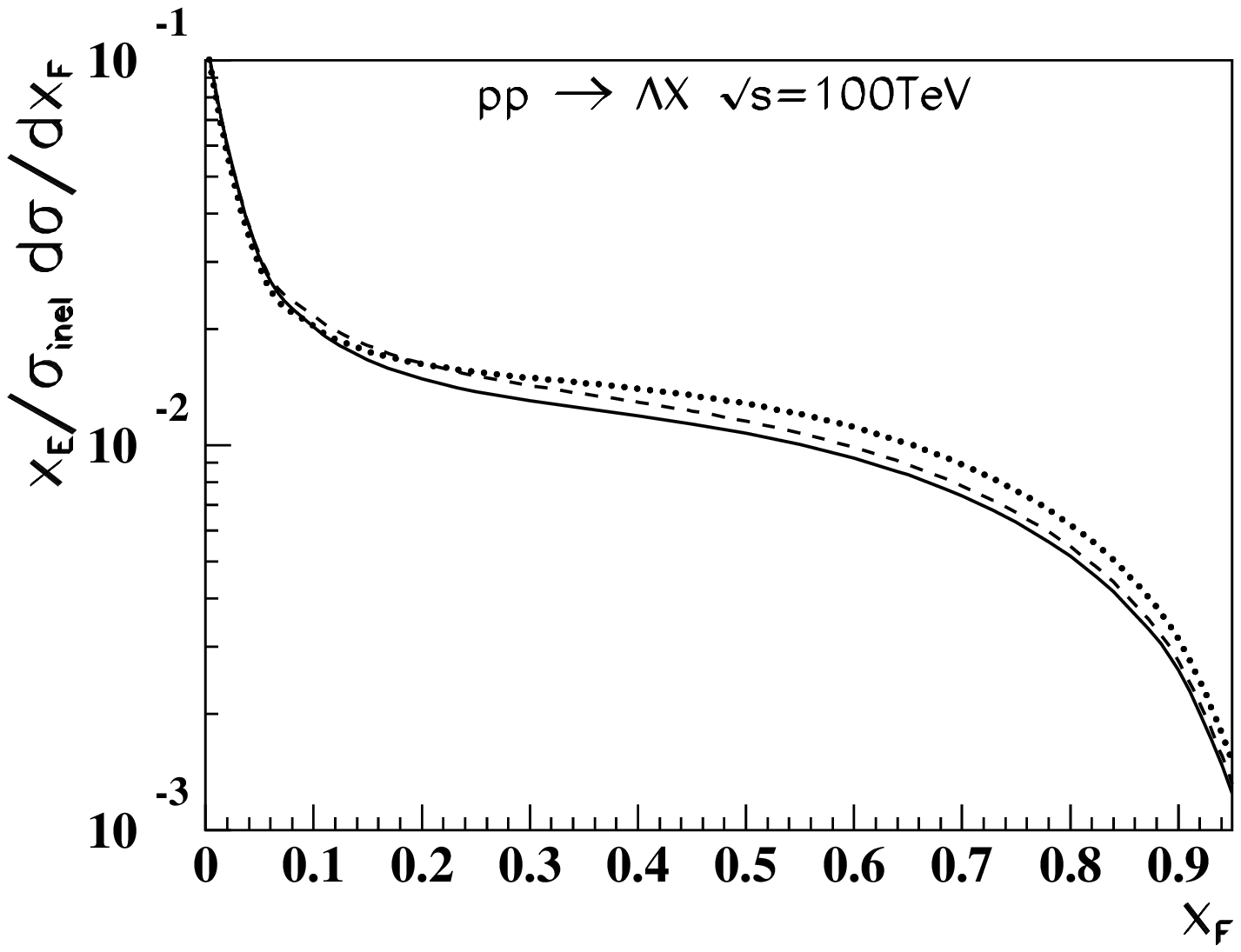}
\vskip -.6cm
\caption{\footnotesize
The QGSM predictions for the spectra of secondary $\Lambda$-hyperons
at energies $\sqrt{s} = 200$ GeV (up left panel), $\sqrt{s} = 900$ GeV
(up right panel), $\sqrt{s} = 7$ TeV (down left panel), and
$\sqrt{s} = 100$ TeV (down right panel). Solid curves correspond to the
value $\alpha_{SJ} = 0.9$, dashed curves to the value $\alpha_{SJ} = 0.5$,
and dotted curves are calculated without SJ contribution ($\varepsilon = 0.$).}
\end{figure}

In the case of secondary $\Lambda$-hyperon production, shown in Fig.~5,
the spectra decrease rather fast at large $x_F$ due to the faster decrease
of $uu$ and $ud$ fragmentation functions into $\Lambda$ in comparison with
their fragmentation into secondary nucleon, and also to the relatively smaller
contribution of diffraction dissociation. The absolute values of $\Lambda$
spectra are smaller than those of the protons and neutrons spectra due to the
strangeness suppression factor.

The Feynman scaling violation effects are shown in more detail in Figs.~6 and
7, for secondary protons and neutrons, respectively. Here, we present the
energy dependences of the spectra at four values of $x_F$, namely at
$x_F = 0.05$, 0.2, 0.5, and 0.7. Separately, we present for the first two
values of $x_F$ (i.e. in the top panels of Figs.~6 and 7) the energy
dependences of net baryon production, i.e. the differences $p-\overline{p}$
and $n-\overline{n}$. In the case of $x_F = 0.5$ and 0.7 the differences
between the baryon
spectra and the net-baryon spectra are negligible.
\begin{figure}[htb]
\centering
\vskip -1.75cm
\includegraphics[width=.49\hsize]{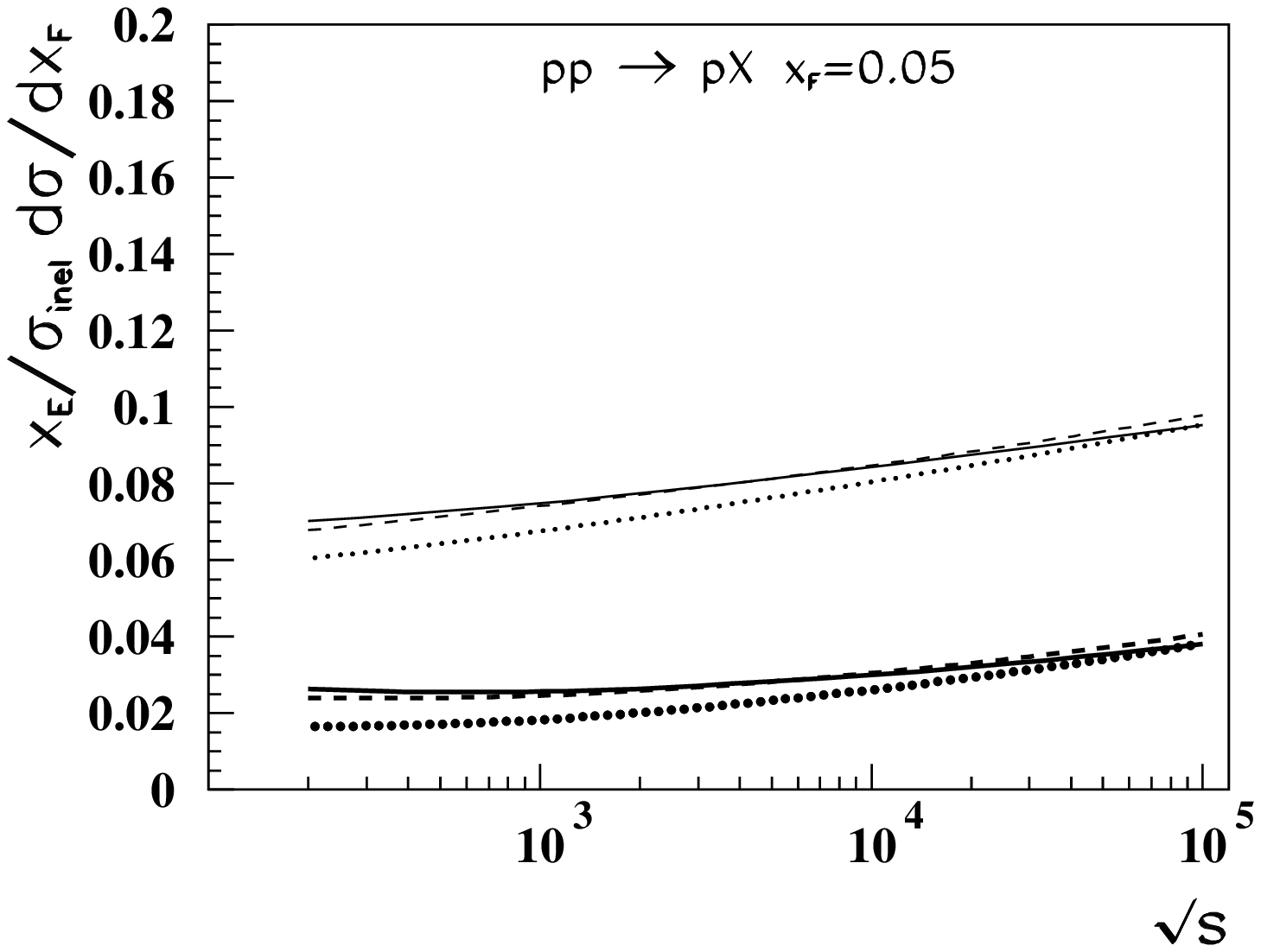}
\includegraphics[width=.49\hsize]{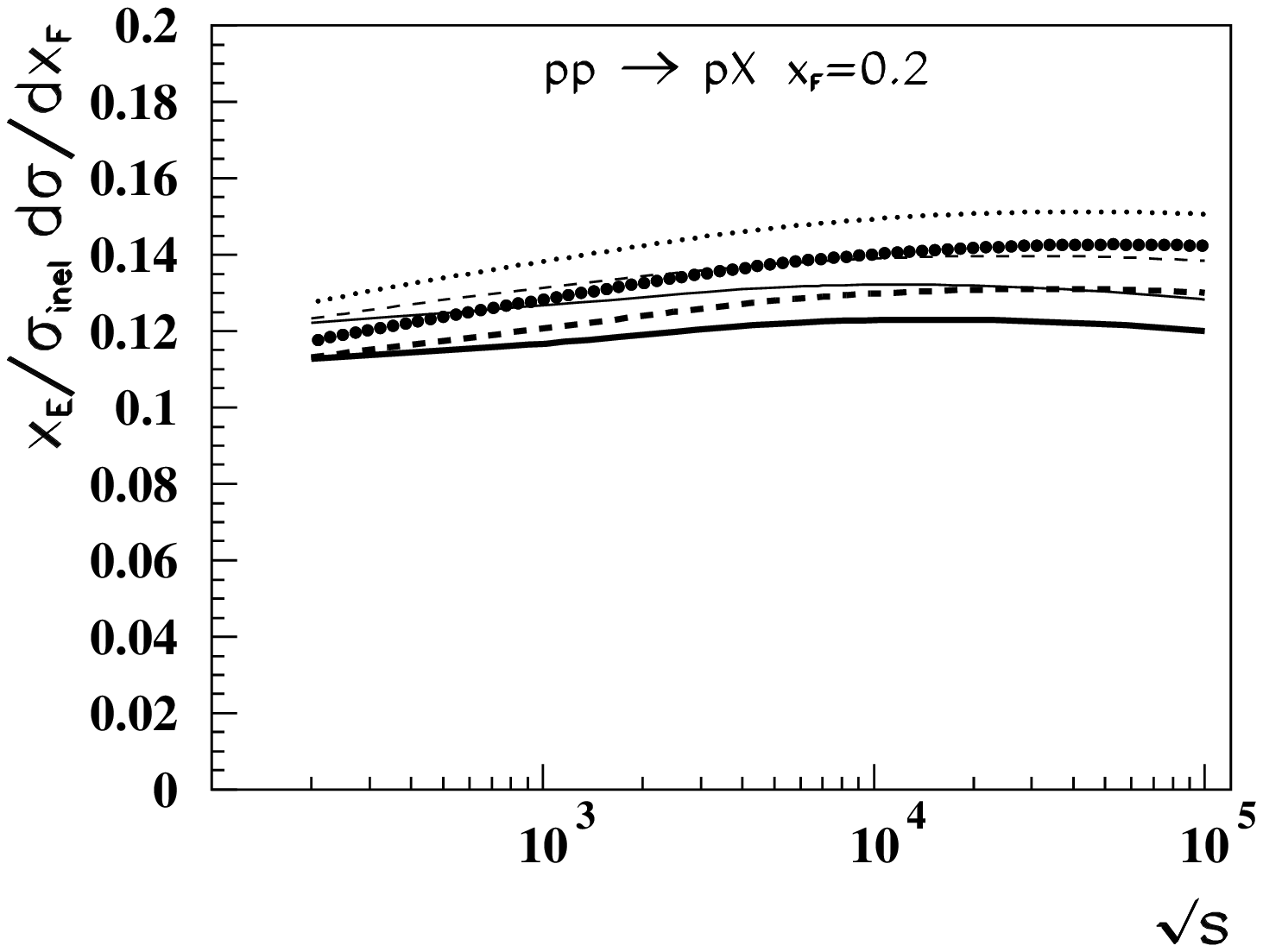}
\vskip -.6cm
\includegraphics[width=.49\hsize]{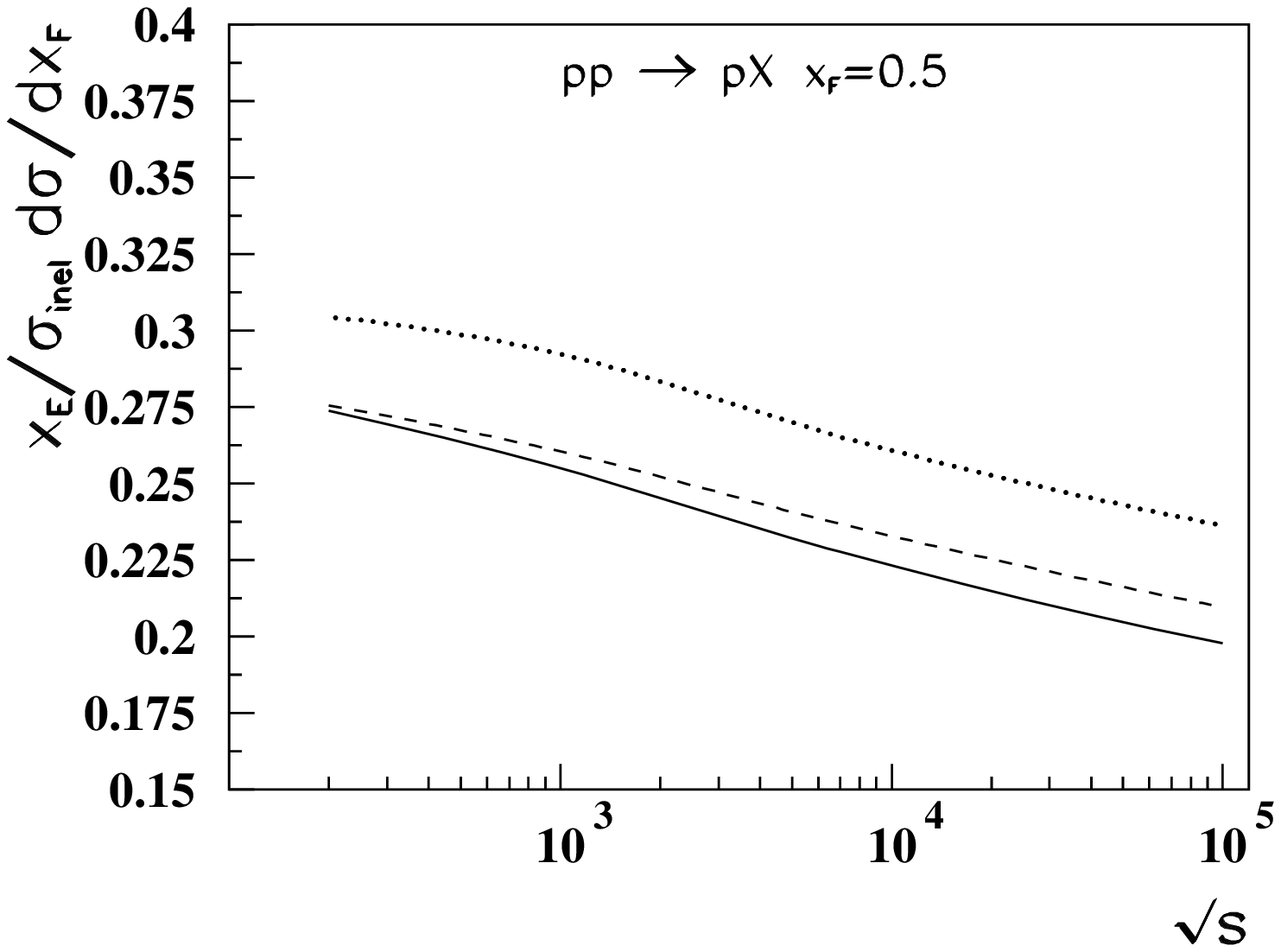}
\includegraphics[width=.49\hsize]{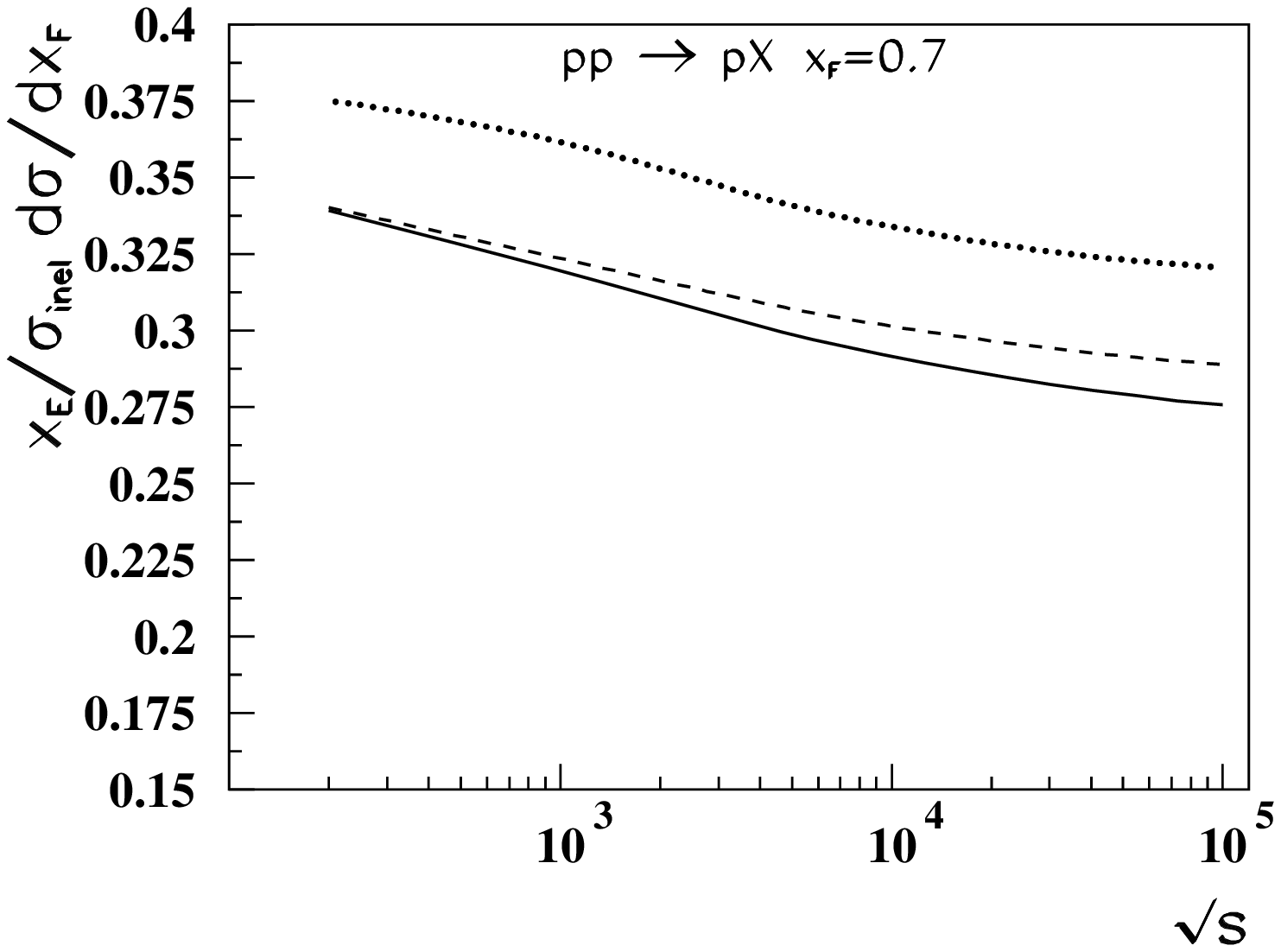}
\vskip -.6cm
\caption{\footnotesize
The QGSM predictions for the spectra of secondary protons as the
functions of energy at fixed values of $x_F$. Thin curves show the total
proton spectra and bold curves at top panels the spectra of net protons,
i.e. the values of the $p-\overline{p}$ differences. Solid curves correspond
to the value $\alpha_{SJ} = 0.9$, dashed curves to the value
$\alpha_{SJ} = 0.5$, and dotted curves are calculated without SJ contribution
($\varepsilon = 0.$).}
\end{figure}

One can see that at $x_F = 0.05$ and 0.2 the spectra of secondary
protons and neutrons increase with energy (both the total spectra as
well as the net baryon spectra). At larger $x_F$ these spectra decrease
with the energy. These energy dependences are connected with the
growth of the average number of exchanging Pomerons. The differences
between solid and dashed curves to dotted curves show the effect of SJ
diffusion.
\begin{figure}[htb]
\centering
\vskip -1.75cm
\includegraphics[width=.49\hsize]{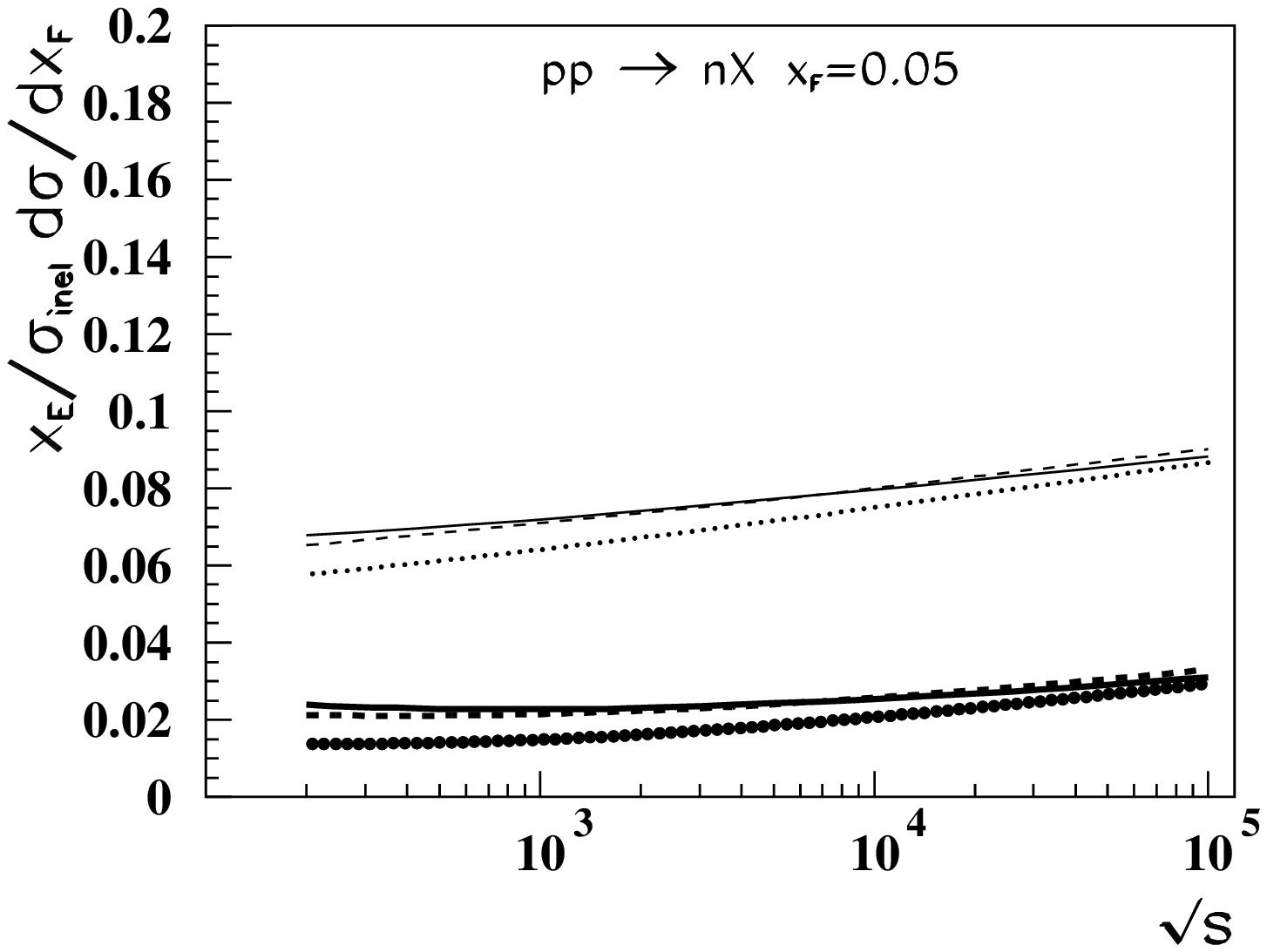}
\includegraphics[width=.49\hsize]{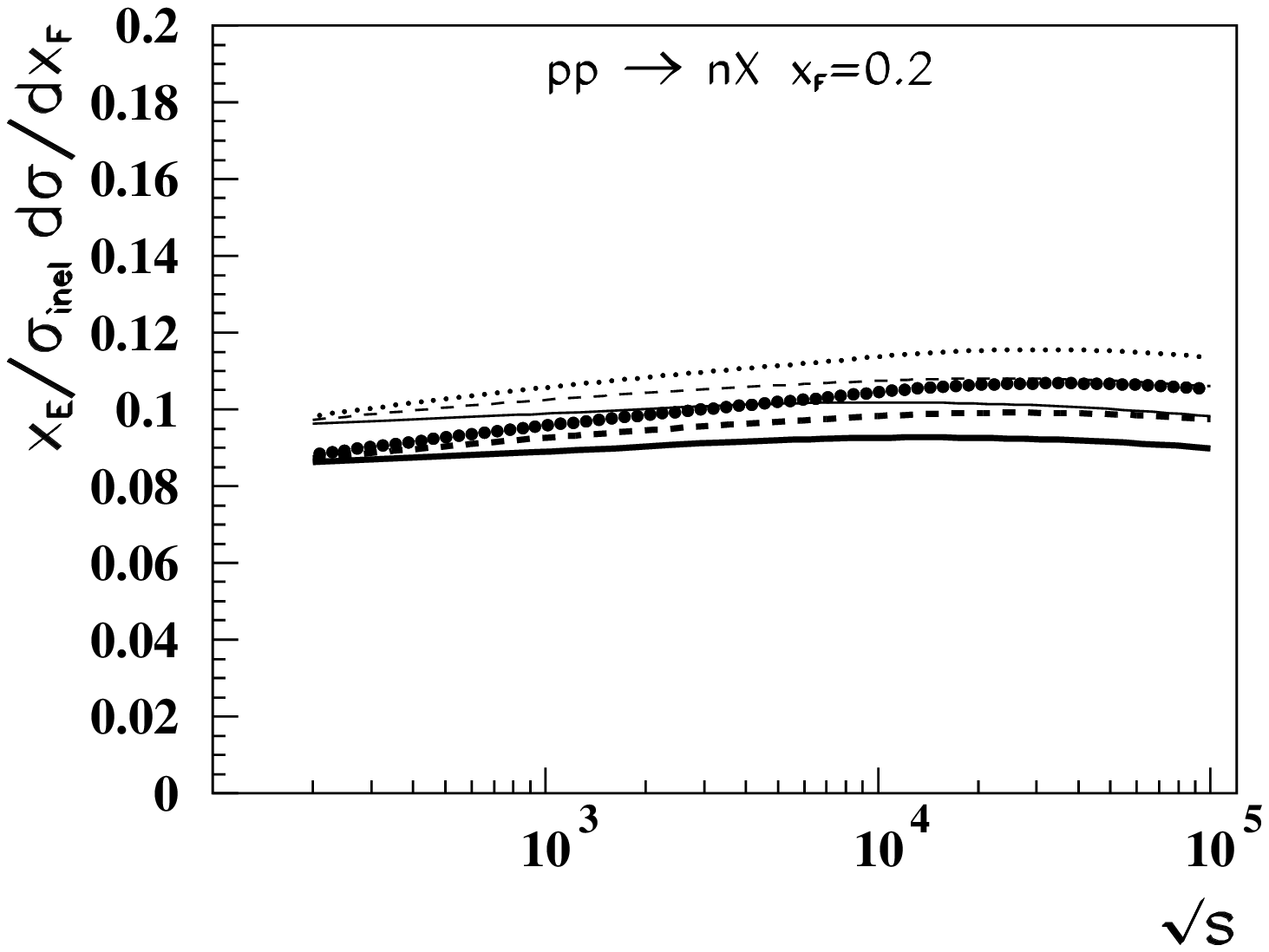}
\vskip -.6cm
\includegraphics[width=.49\hsize]{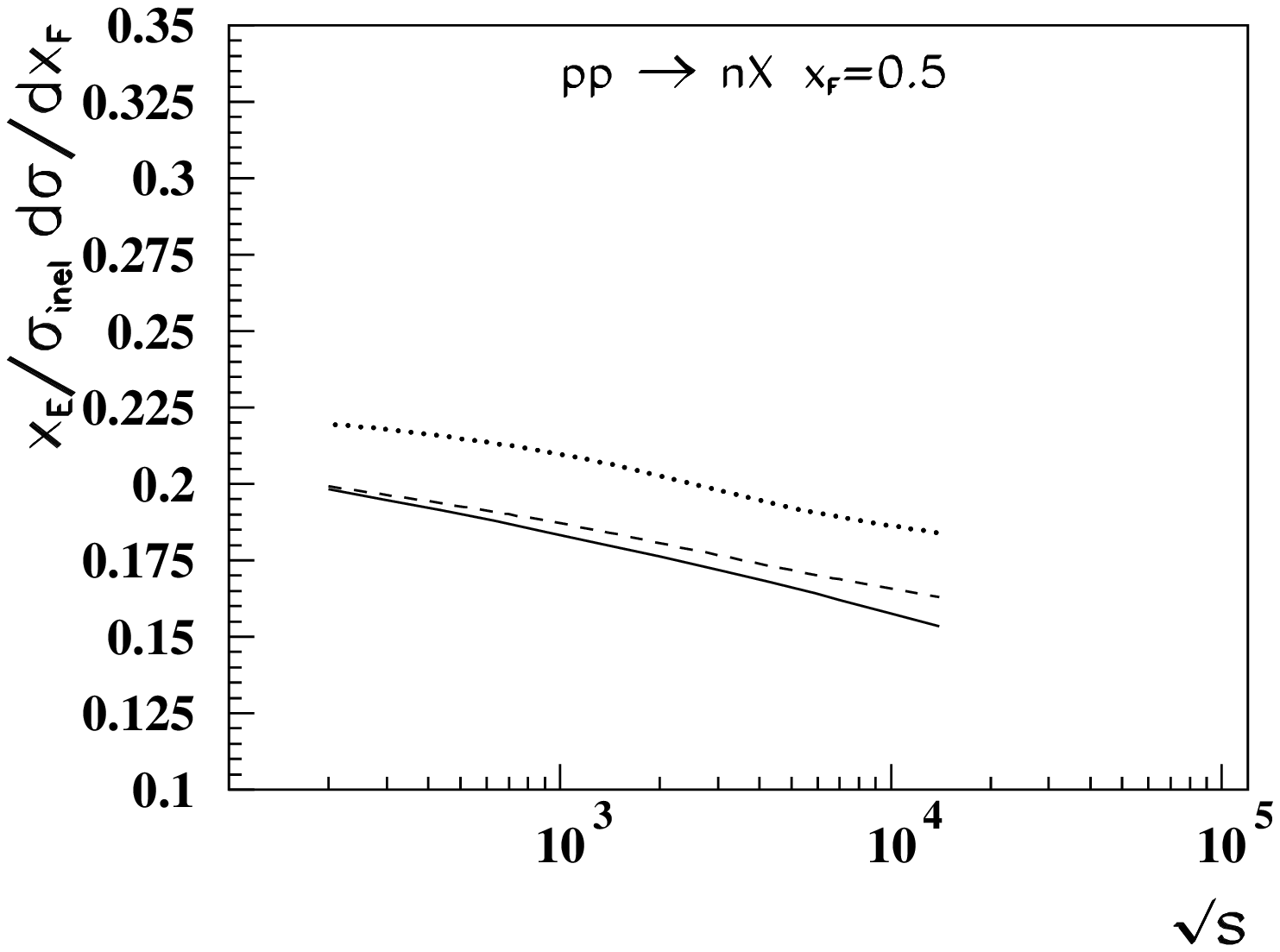}
\includegraphics[width=.49\hsize]{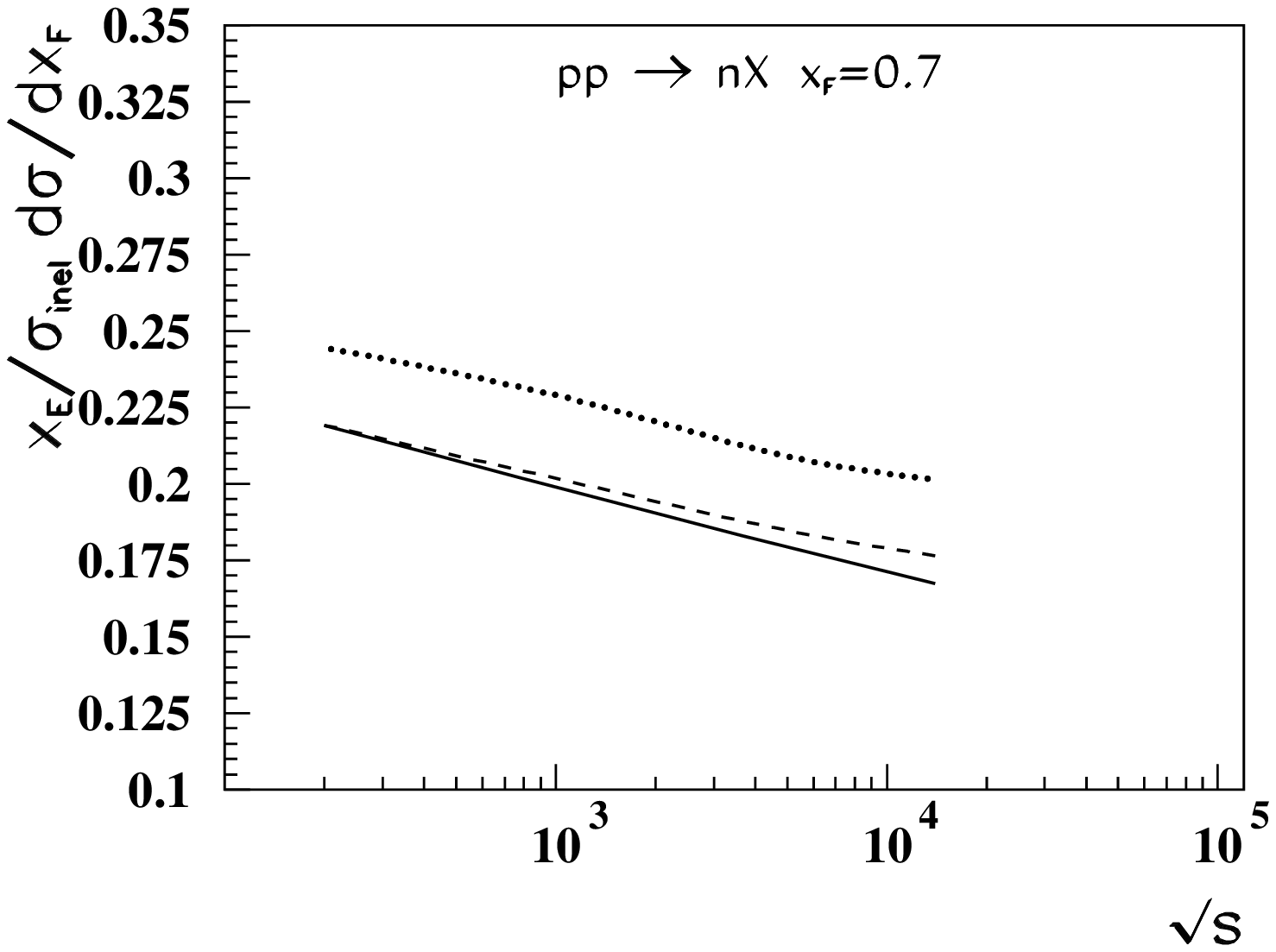}
\vskip -.6cm
\caption{\footnotesize
The QGSM predictions for the spectra of secondary neutrons as the
functions of energy at fixed values of $x_F$. Thin curves show the total
neutron spectra and bold curves at top panels the spectra of net neutrons,
i.e. the values of $n-\overline{n}$ differences. Solid curves correspond to the
value$ \alpha_{SJ} = 0.9$, dashed curves to the value $\alpha_{SJ} = 0.5$,
and dotted curves are calculated without SJ contribution ($\varepsilon = 0.$).}
\end{figure}

\section{The ratios of the inclusive spectra of baryons at different
energies}

The possible Feynman scaling violation at superhigh energies was
discussed in \cite{LHCf1}, based on Monte Carlo calculations. These
effects are, as a rule, numerically not large, and so it is more suitable
to consider the ratios of the spectra at different energies. These
ratios calculated in the QGSM at $\sqrt{s} = 7$ TeV, 14 TeV, and 1000
TeV to the values at $\sqrt{s} = 900$ GeV, are presented both for
secondary protons and neutrons in Fig.~8.

In Fig.~9 we present the ratios of the spectra of secondary protons
and neutrons as the functions of rapidity, calculated with
$\alpha_{SJ} = 0.9$ and without SJ contribution, at $\sqrt{s} = 200$
GeV (solid curves) and at $\sqrt{s} = 100$ TeV (dashed curves).
The similar ratios calculated with $\alpha_{SJ} = 0.5$ and without
SJ contribution are shown by dash-dotted curves at $\sqrt{s} = 200$
GeV and by dotted curves at $\sqrt{s} = 100$ TeV.

\section{Conclusion}

We present the QGSM predictions for Feynman scaling violation in
the spectra of leading baryons due to baryon charge diffusion at
large distances in the rapidity space. The existance
of such a diffusion has been attested observed in many papers, even
at the LHC energies, and so the decrease of the spectra in the fragmentation
region is the direct consequence of baryon charge conservation.
However, the numerical values of the scaling violation at
different $x_F$ are model-dependent. On the other point of view,
the calculations of scaling violation for leading baryons should
be accompanied by the calculation of the baryon/antibaryon
asymmetry in the central region.

The first experimental data obtained at LHC are in general agreement with
the QGSM calculations performed with the same values of parameters which were
determined at lower energies (mainly for the description of the the fixed
target experiments). However, the numerical value of $\alpha_{SJ}$ is not
well-known. ALICE Collaboration data on the ratios of $\overline{p}p$
production are in agreement with a value $\alpha_{SJ} = 0.5$, while LHCb
Collaboration data on the ratios of $\overline{\Lambda}\Lambda$ production do
not allow to fix the value of $\alpha_{SJ}$, and there are two experimental
results for $\overline{B}B$ production asymmetry.

\begin{figure}[htb]
\centering
\vskip -1.cm
\vskip -.4cm
\includegraphics[width=.49\hsize]{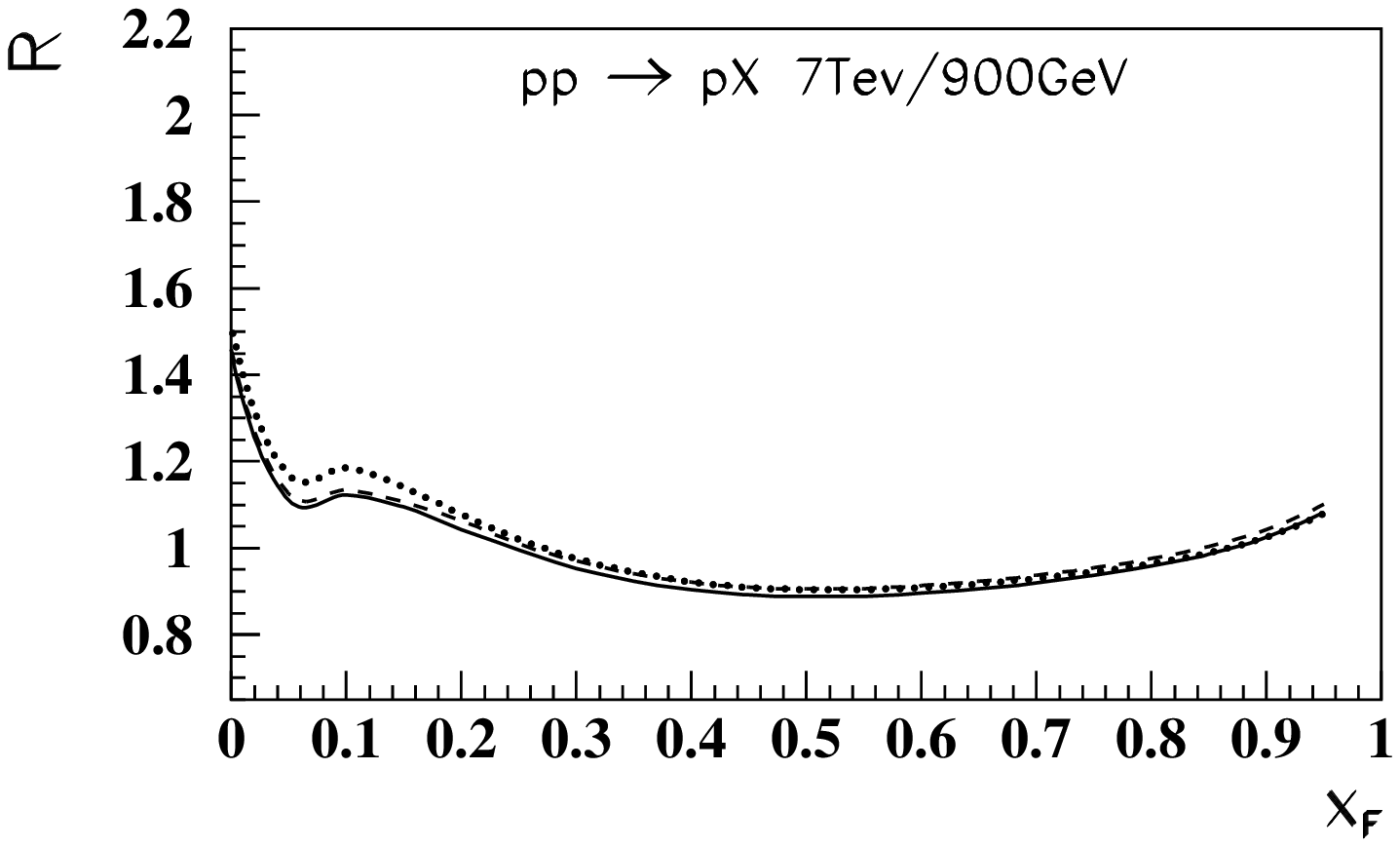}
\includegraphics[width=.49\hsize]{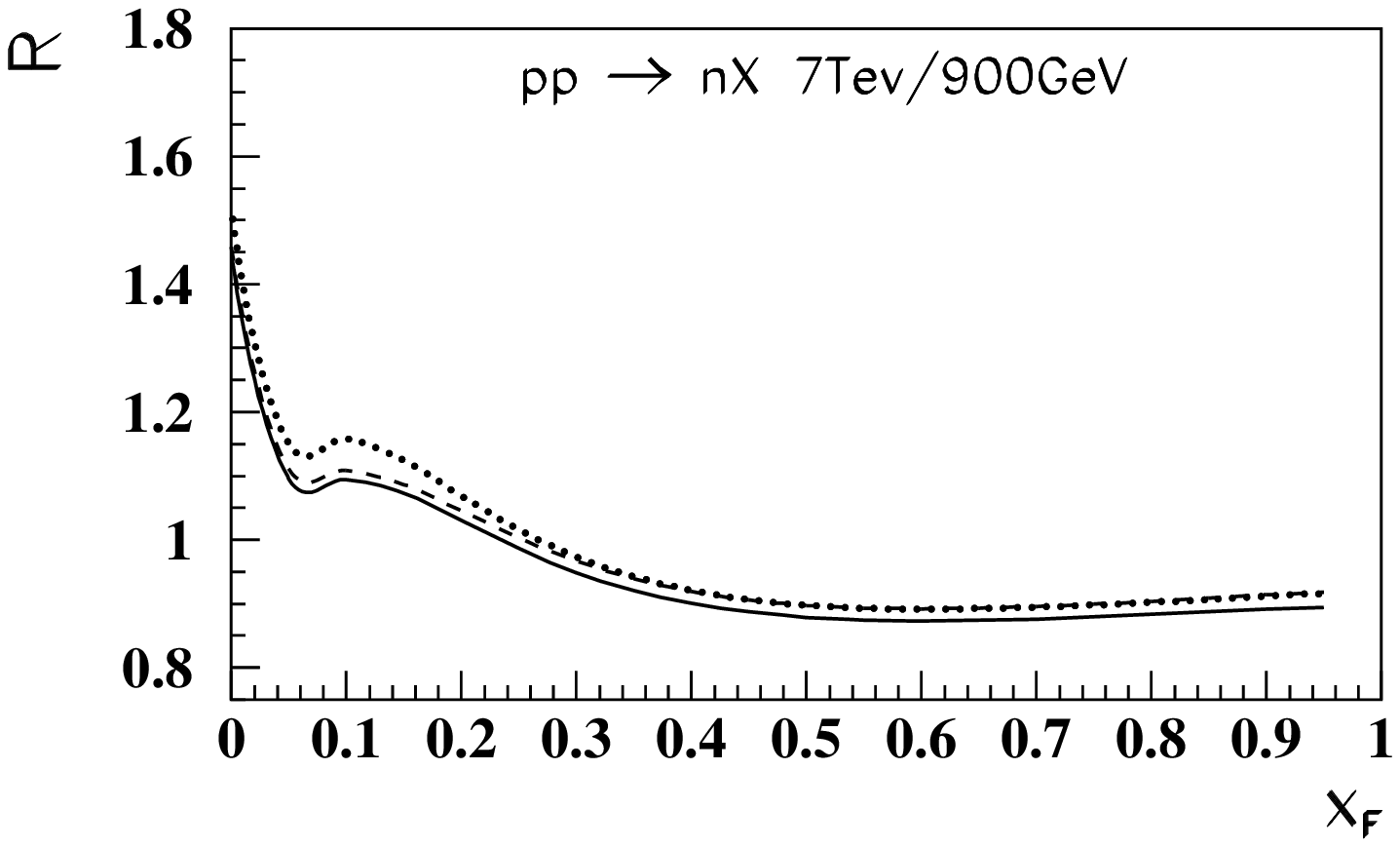}
\vskip -.8cm
\includegraphics[width=.49\hsize]{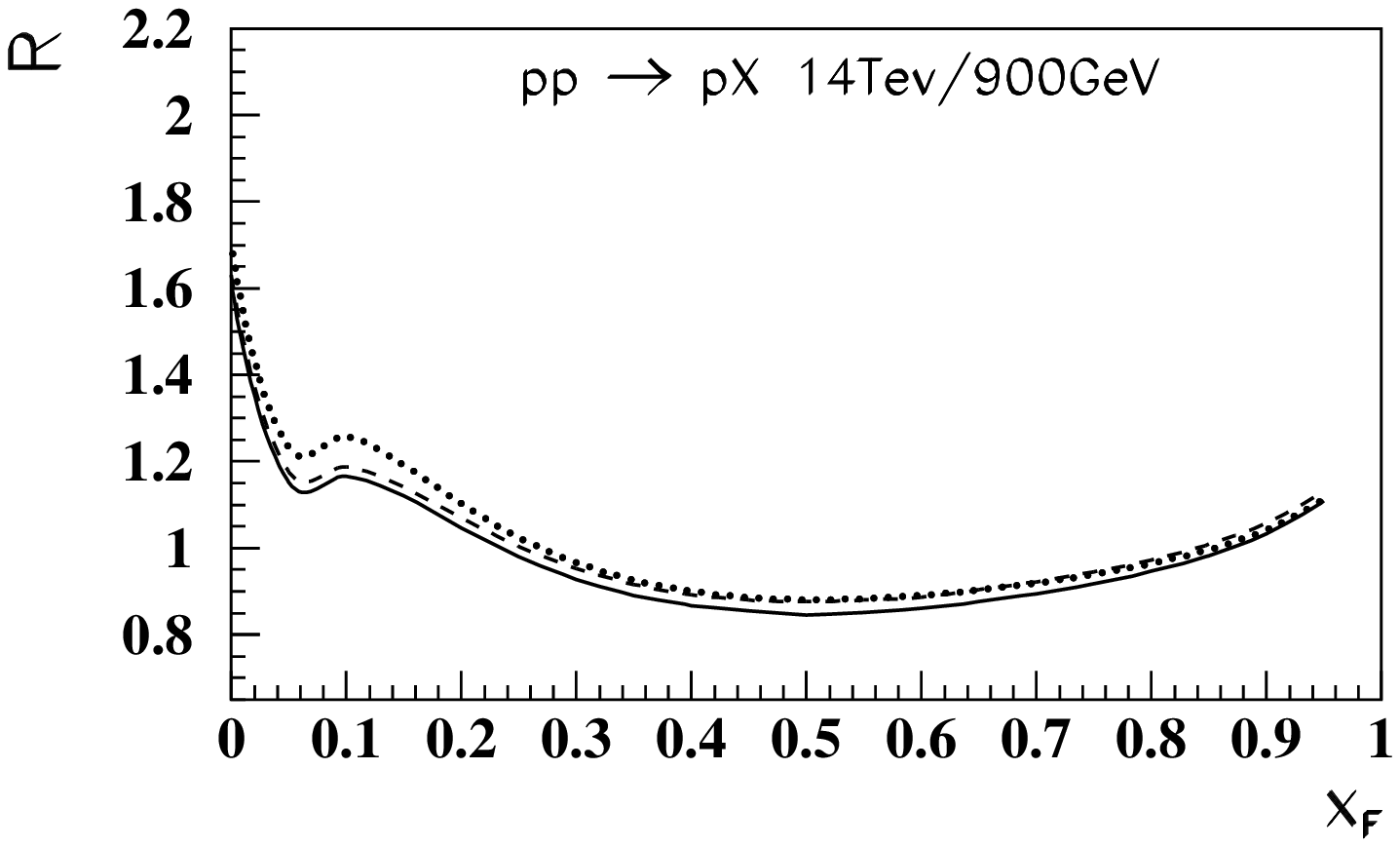}
\includegraphics[width=.49\hsize]{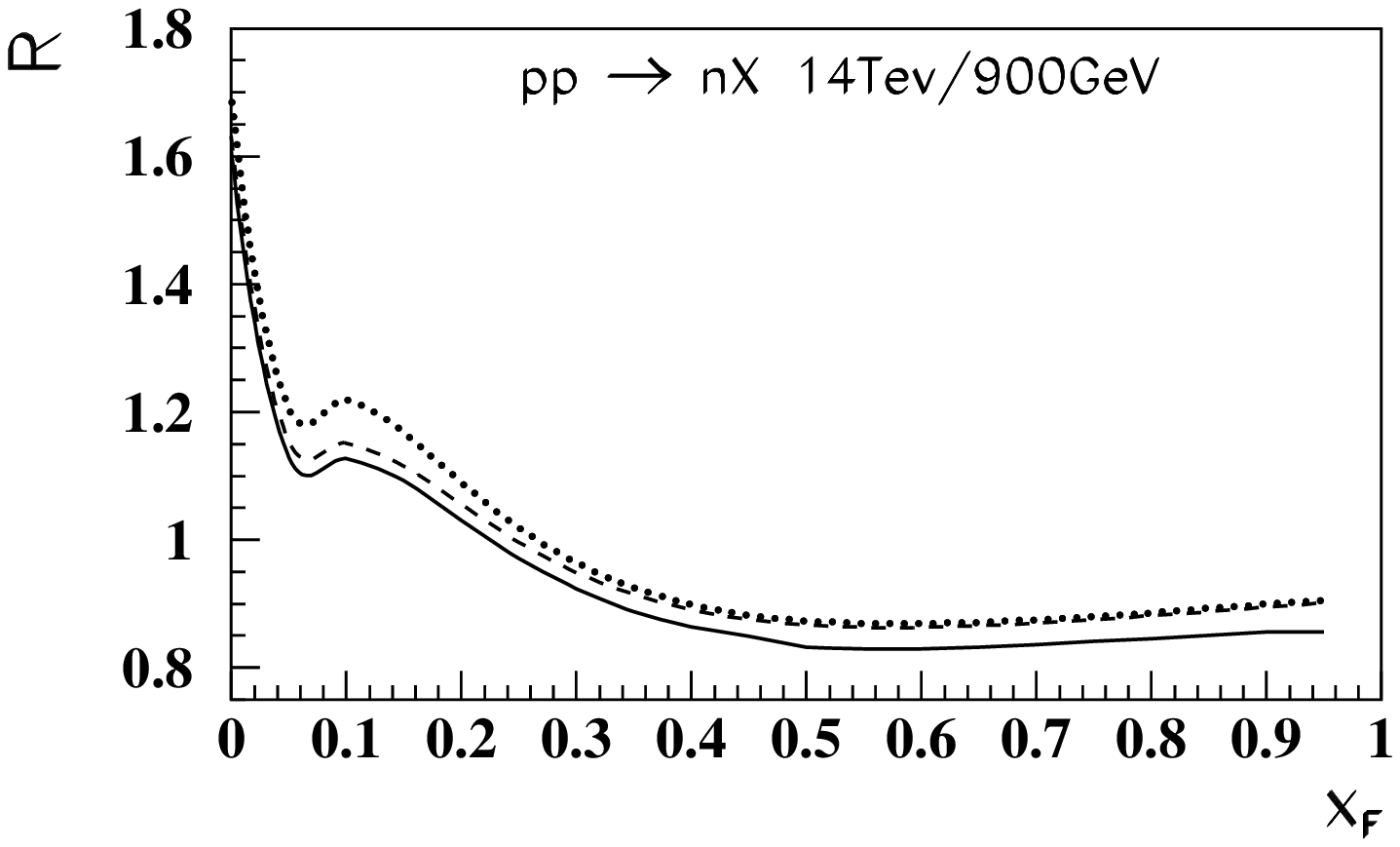}
\vskip -.8cm
\includegraphics[width=.49\hsize]{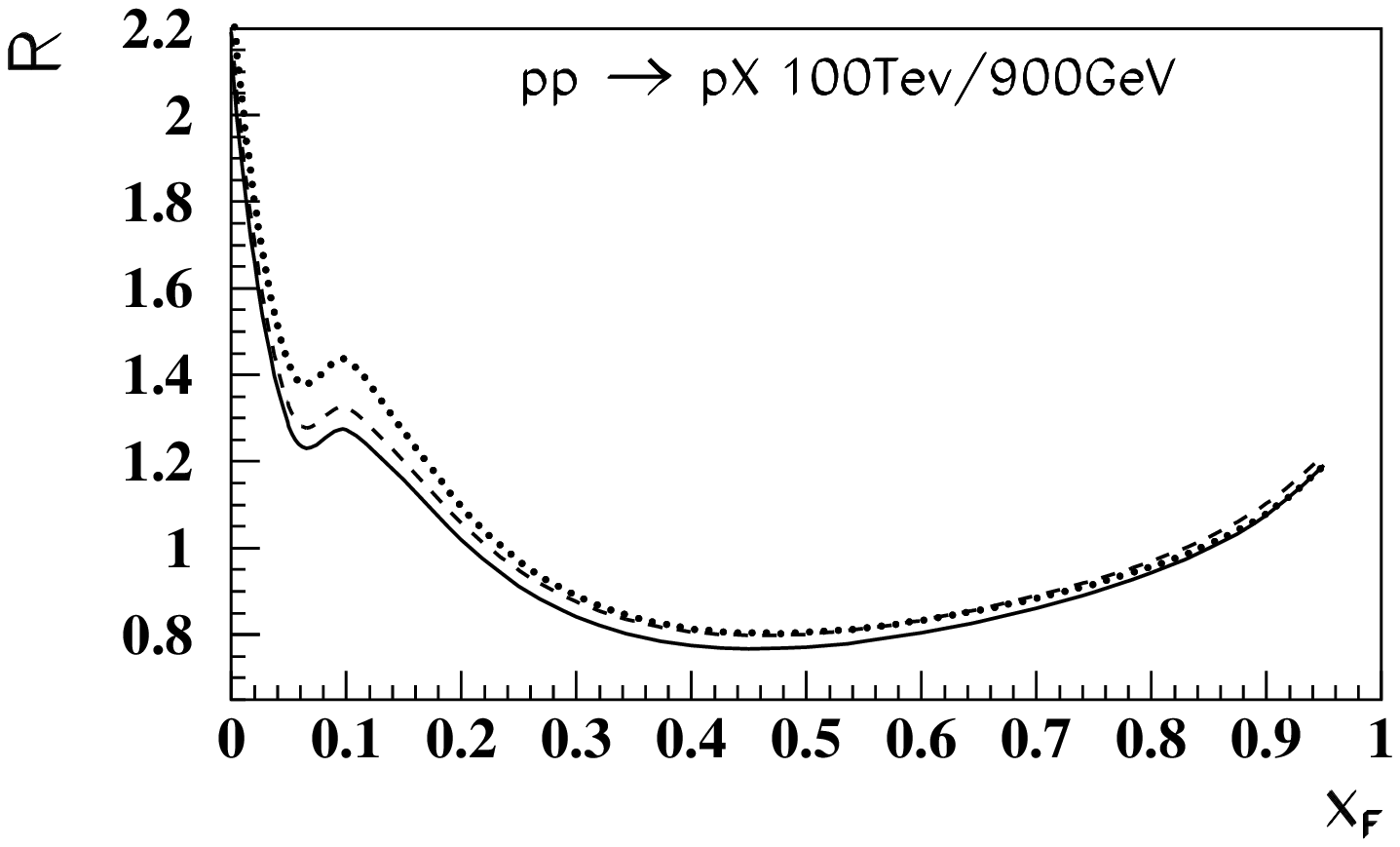}
\includegraphics[width=.49\hsize]{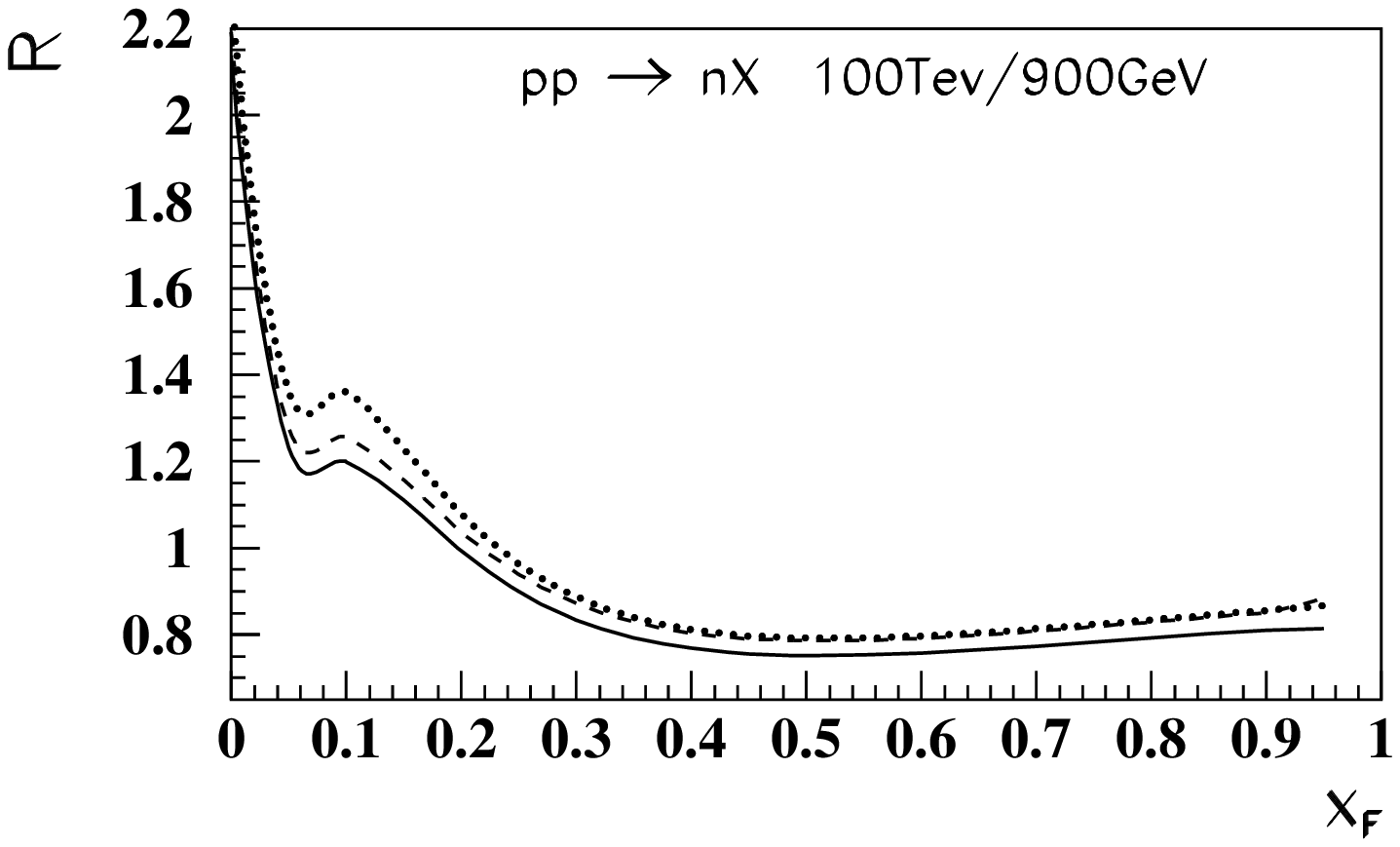}
\vskip -.4cm
\caption{\footnotesize
The QGSM predictions for the ratios of the spectra of secondary protons
(left panels) and neutrons (right panels) at three different energies to
those at $\sqrt{s} = 900$ GeV.}
\end{figure}

We neglect by the possibility of interactions between Pomerons the (so-called
enhancement diagrams), since our estimations \cite{MPS} show that the inclusive
density of secondaries produced in $pp$ collisions at LHC energies is not large
enough for these diagrams to be significant.

Our calculations are in reasonable agreement with the results of
ref. \cite{BBKZ}.

\begin{figure}[htb]
\centering
\includegraphics[width=.49\hsize]{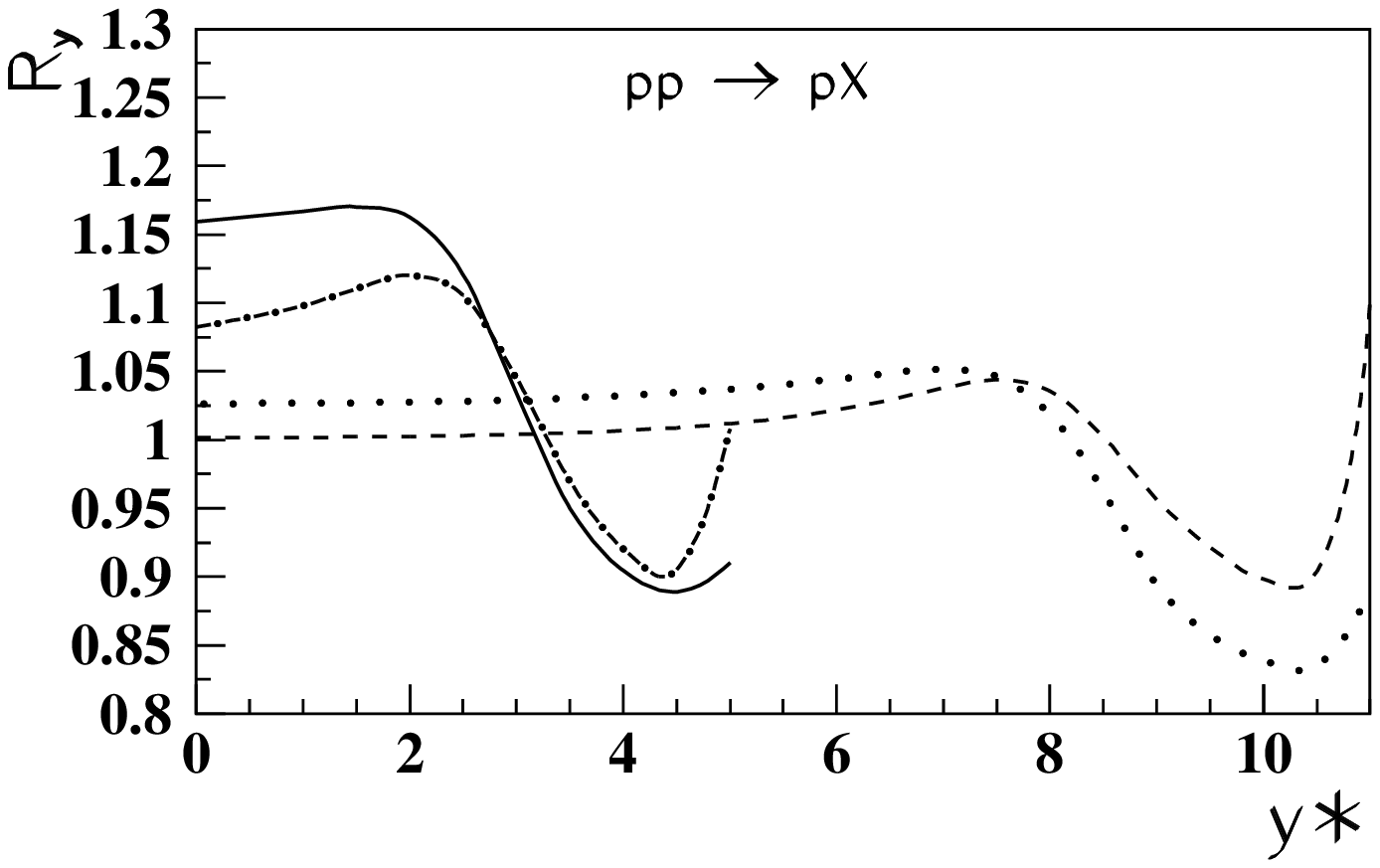}
\includegraphics[width=.49\hsize]{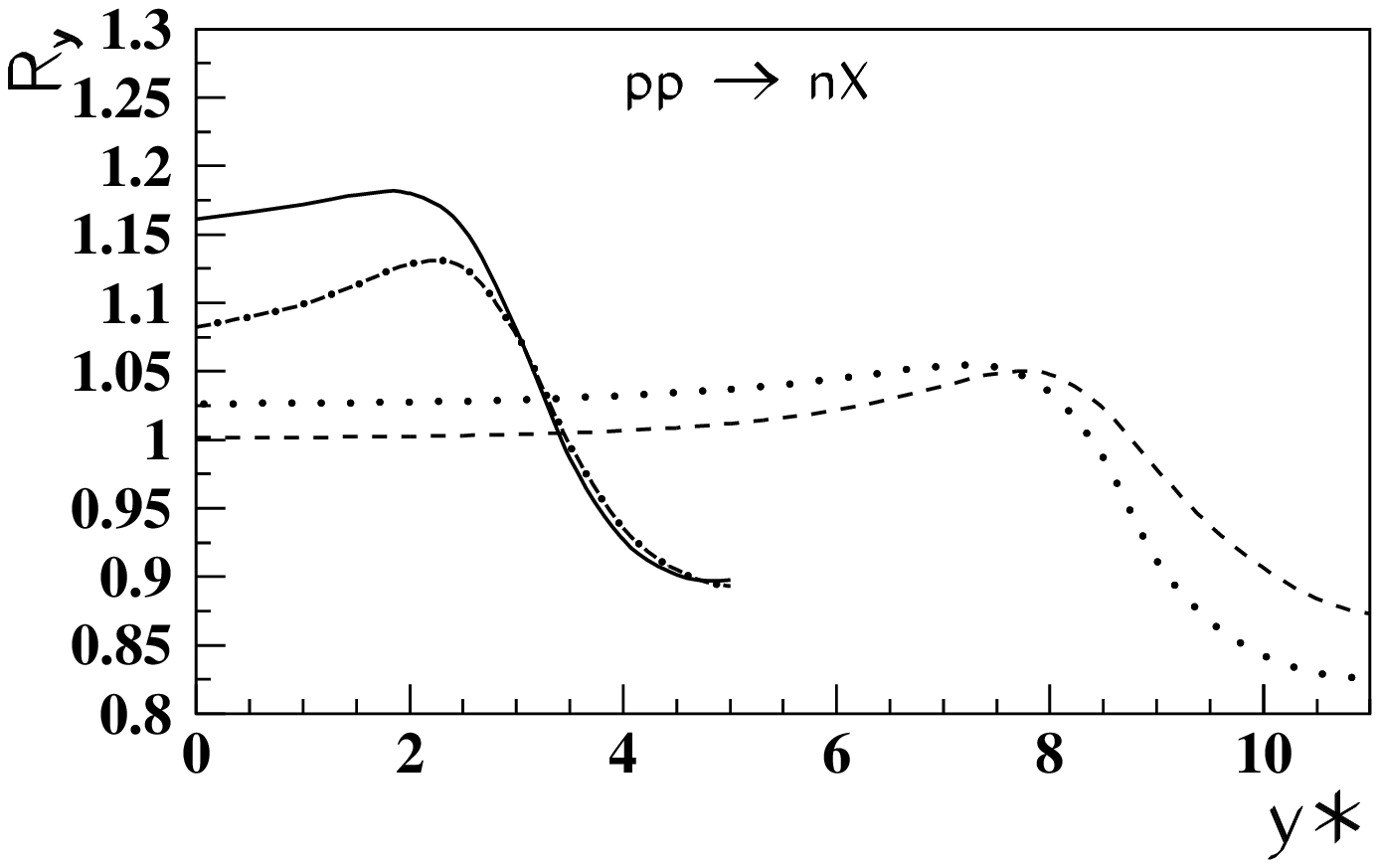}
\vskip -.4cm
\caption{\footnotesize
The QGSM predictions for the ratios of the spectra of secondary protons
(left panels) and neutrons (right panels) calculated with $\alpha_{SJ} = 0.9$
(solid and dashed curves) and with $\alpha_{SJ} = 0.9$ (dash-dotted and
dotted curves) and with energies $\sqrt{s} = 200$ GeV (solid and dash-dotted
curves) and at $\sqrt{s} = 100$ TeV (dashed and dotted curves).}
\end{figure}

{\bf Acknowledgements}

We are grateful to \frame{A.B. Kaidalov} for useful discussions
and comments. This paper was supported by Ministerio de Educaci\'on y
Ciencia of Spain under the Spanish Consolider-Ingenio 2010 Programme
CPAN (CSD2007-00042) and project FPA 2005--01963, by Xunta de Galicia, Spain,
and, in part, by grant RFBR 11-02-00120-a and by the Department of Education
and Science of the Republic of Armenia, Grant-11-1C015.


\newpage

\begin{thebibliography}{**}

\bibitem{Abr} J. Abraham et al., Phys. Rev. Lett. {\bf 104}, 091101 (2010)
and arXiv:1002.0699 [astro-ph.HE].

\bibitem{Abb} R.U. Abbasi et al., Phys. Rev. Lett. {\bf 104}, 161101 (2010)
and arXiv:0910.4184 [astro-ph.HE].

\bibitem{ELF} E.L. Feinberg, Phys. Rep. {\bf 50}, 237 (1972).

\bibitem{Ver} S.N. Vernov et al., J. Phys {\bf C3}, 1601 (1977).

 \bibitem{KKh} M.N. Kalmykov and G.B. Khistiansen, Jetp. Lett. {\bf37},
247 (1983).

\bibitem{LHCf1} T. Sako, LHCf Collaboration, arXiv:1010.0195 [hep-ex].

\bibitem{LHCf2} O. Adriani et al., LHCf Collaboration, arXiv:1012.1490
[hep-ex].


\bibitem{ASS} V.V.~Anisovich, Yu.M. Shabelski, and V.M.~Shekhter,
Nucl. Phys. B~{\bf 5133}, 477 (1978).

\bibitem{ABS} V.V.~Anisovich, V.M.~Braun, and Yu.M. Shabelski,
Z. Phys. C~{\bf 27}, 77 (1985).

\bibitem{KTM} A.B. Kaidalov and K.A. Ter-Martirosyan, Yad. Fiz. {\bf 39},
1545 (1984); {\bf 40}, 211 (1984).

\bibitem{KaPi} A.B. Kaidalov and O.I.~Piskounova, Yad. Fiz. {\bf 41}, 1278
(1985); Z. Phys. {\bf C30}, 145 (1986).

\bibitem{Sh} Yu.M. Shabelski, Yad. Fiz. {\bf 44}, 186 (1986).

\bibitem{Ans} F. Anselmino, L. Cifarelli, E. Eskut, and Yu.M. Shabelski,
Nouvo Cim. {\bf 105A}, 1371 (1992).

\bibitem{ACKS} G.H. Arakelyan, A. Capella, A.B.~Kaidalov, and
Yu.M.~Shabelski, Eur. Phys. J. C~{\bf26}, 81 (2002) and hep-ph/0103337.

\bibitem{KTMS} A.B. Kaidalov, K.A. Ter-Martirosyan, and Yu.M.~Shabelski,
Yad. Fiz. {\bf 43}, 1282 (1986).

\bibitem{Sh1} Yu.M. Shabelski, Z. Phys. {\bf C38}, 569 (1988).

\bibitem{JDDS} J. Dias de Deus and Yu.M. Shabelski, Yad. Fiz. {\bf 71}, 191
(2008).

\bibitem{AGK} V.A. Abramovsky, V.N. Gribov, and O.V.~Kancheli, Yad. Fiz.
{\bf 18}, 595 (1973).

\bibitem{Kai} A.B. Kaidalov, Sov. J. Nucl. Phys. {\bf 45}, 902 (1987);
Yad. Fiz. {\bf 43}, 1282 (1986).

\bibitem{MPRS} C. Merino, C. Pajares, M.M. Ryzhinskiy, and Yu.M.~Shabelski, arXiv:1007.3206[hep-ph].

\bibitem{Sh2} Yu.M. Shabelski, Yad. Fiz. {\bf 45}, 223 (1987).

\bibitem{Sh3} Yu.M. Shabelski, Z. Phys. {\bf C38}, 569 (1988).

\bibitem{EKS} A.D. Erlykin, N.P. Krutikova, and Yu.M. Shabelski,
Yad. Fiz. {\bf 45} 1075 (1987), {\bf 47} 1667 (1988).

\bibitem{BS} F. Bopp and Yu.M. Shabelski, Yad. Fiz. {\bf 68}, 2155 (2005)
and hep-ph/0406158; Eur. Phys. J. A~{\bf 28}, 237 (2006) and hep-ph/0603193.

\bibitem{AMPS} G.H. Arakelyan, C. Merino, C. Pajares, and Yu.M.~Shabelski,
Eur. Phys. J. {\bf C54}, 577 (2008) and hep-ph/0709.3174.

\bibitem{MRS1} C. Merino, M.M. Ryzhinskiy, and Yu.M.~Shabelski, Eur. Phys. J.
{\bf C62}, 491 (2009); arXiv:0810.1275 [hep-ph].

\bibitem{AKMS} G.H. Arakelyan, A.B. Kaidalov, C. Merino, and Yu.M.~Shabelski,
Phys. Atom. Nucl. {\bf 74}, 426 (2011) and arXiv:1004.4074 [hep-ph].

\bibitem{MPS1} C. Merino, C. Pajares, and Yu.M.~Shabelski,
Eur. Phys. J. {\bf C71}, 1652 (2011); arXiv:1105.3174 [hep-ph].




\bibitem{Artru} X. Artru, Nucl. Phys. B {\bf85}, 442 (1975).

\bibitem{IOT} M. Imachi, S. Otsuki, and F.~Toyoda, Prog. Theor. Phys.
{\bf 52}, 346 (1974); {\bf 54}, 280 (1976); {\bf 55}, 551 (1976).

\bibitem{RV} G.C. Rossi and G.~Veneziano, Nucl. Phys. B~{\bf 123}, 507 (1977).

\bibitem{Khar} D. Kharzeev, Phys. Lett. B~{\bf 378}, 238 (1996).

\bibitem{latt} V.G. Bornyanov et al., Uspekhi Fiz. Nauk. {\bf 174}, 19 (2004).

\bibitem{AMS} G.H. Arakelyan, C. Merino, and Yu.M.~Shabelski, Yad. Fiz.
{\bf 69}, 911 (2006) and hep-ph/0505100; Phys. Atom. Nucl. {\bf 70}, 1110
(2007) and hep-ph/0604103; Eur. Phys. J. {\bf A31}, 519 (2007) and hep-ph/0610264.

\bibitem{Olga} O.I. Piskounova, Phys. Atom. Nucl. {\bf 70}, 1110 (2007) and
hep-ph/0604157.

\bibitem{MRS} C. Merino, M.M. Ryzhinskiy, and Yu.M.~Shabelski, Proceedings of the
XLIII PNPI Winter School on Nuclear and Particle Physics (PNPI-2009), Repino,
St.Petersburg, Russia, February 24$^{th}$-March 1$^{st}$, 2009, pages 156-185,
and arXiv:0906.2659 [hep-ph].


\bibitem{VGW} S.E. Vance, M. Gyulassy, and X-N. Wang, Phys. Lett. B~{\bf 443},
45 (1998).

\bibitem{ait} E.M. Aitala {\em et al.}, E769 Collaboration, hep-ex/0009016;
Phys. Lett. B~{\bf469}, 9 (2000).

\bibitem{KP1} B.Z. Kopeliovich and B.~Povh, Z. Phys. C~{\bf 75}, 693 (1997);
Phys. Lett. B~{\bf 446}, 321 (1999).




\bibitem{AnSh} V.V.~Anisovich and V.M.~Shekhter, Nucl. Phys. B~{\bf 55},
455 (1973).

\bibitem{CS} A. Capella and C.A. Salgado, Phys. Rev. C~{\bf 60}, 054906 (1999).


\bibitem{ALICE} K. Aamodt et al., ALICE Collaboration,
Phys. Rev. Lett. {\bf105}, 072002 (2010) and arXiv:1006.5432 [hep-ex].

\bibitem{LHCb} F. Dettori, LHCb Collaboration, hep-ex/1009.1221.

\bibitem{Conf} C. Merino, C. Pajares, and Yu.M.~Shabelski, Proceedings of the
International Conference on Hadron Structure and QCD, Gatchina (2010), to appear.

\bibitem{ZEU1} S. Chekanov et al., ZEUS Collaboration, Nucl. Phys. {\bf B658}, 3 (2003) and hep-ex/0210029.

\bibitem{ZEU2} S. Chekanov et al., ZEUS Collaboration, JHEP {\bf 0906}, 74 (2009) and arXiv:0812.2416 [hep-ex].

\bibitem{ZEU3} S. Chekanov et al., ZEUS Collaboration, Nucl. Phys. {\bf B776}, 1 (2007) and hep-ex/0702028.


\bibitem{KKMR} A B. Kaidalov et al., Eur. Phys. J. {\bf C47}, 385 (2006).

\bibitem{KPSS} B.Z. Kopeliovich et al., Phys. Rev. {\bf D78}, 014031
(2008) and arXiv:0805.4534 [hep-ph].



\bibitem{MPS} C. Merino, C. Pajares, and Yu.M.~Shabelski, Eur. Phys. J.
{\bf C59}, 691 (2009) and arXiv:0802.2195 [hep-ph].

\bibitem{BBKZ} C.J. Bleibel et al., arXiv:1011.2703 [hep-ex].

\end{thebibliography}
\end{document}